\documentclass[11pt,notitlepage,nofootinbib]{revtex4-1}

\usepackage{epsf}
\usepackage{latexsym}
\usepackage{amssymb}
\usepackage{amsmath}
\usepackage{bm}
\usepackage[dvips]{graphicx}
\usepackage{array}

\date{\today}
\def\d3{^{(3)}\nabla}

\begin{document}

\title{21 cm line signal from magnetic modes}

\author{Kerstin E. Kunze}

\email{kkunze@usal.es}

\affiliation{Departamento de F\'\i sica Fundamental, Universidad de Salamanca,
 Plaza de la Merced s/n, 37008 Salamanca, Spain}

\begin{abstract}
The Lorentz term raises the  linear matter power on small scale which leads to interesting signatures in the 21 cm signal. Numerical simulations of the resuting nonlinear density field, the distribution of ionized hydrogen and the 21 cm signal at different values of redshift are presented for
magnetic fields with field strength B=5 nG, and spectral indices $n_B=-2.9, -2.2$ and -1.5 together with the adiabatic mode for the best fit data  of Planck13+WP. Comparing the averaged global 21 cm signal with the projected SKA1-LOW sensitivities of the Square Kilometre Array (SKA) it might be possible to constrain the magnetic field parameters.  
\end{abstract}

\maketitle

\section{Introduction}
\label{s0}
\setcounter{equation}{0}

Magnetic fields come in different shapes and sizes in the universe. Observed magnetic fields  range from those  associated with stars and planets upto to cluster and super cluster scales (cf., e.g., \cite{Govoni:2017qmd, Pfrommer:2012mm, Han:2002ns}). Observations of the energy spectra of 
a number of blazars in the GeV range with  Fermi/LAT and  in the TeV range with telescopes such as  H.E.S.S., MAGIC or VERITAS  
have been interpreted as evidence for the existence of truly cosmological magnetic fields. These are not associated with virialized structures but rather
permeate the universe. Limits on the field strengths of  these void magnetic fields are of the order $10^{-25}-10^{-15}$ G (e.g, \cite{Takahashi:2013lba, Essey:2010nd, Tavecchio:2010mk})which is considerably below those of galactic magnetic fields which are of the order of $10^{-6} $ G (e.g., \cite{Boulanger:2018zrk}).

Cosmological magnetic fields present from before decoupling influence the cosmic plasma in different ways. Before recombination photons are strongly coupled to the baryon fluid via Thomson scattering off the electrons. As the observed high degree of isotropy on large scales limits the magnitude of a homogeneous magnetic field the contribution of a putative, primordial magnetic field is modeled as a gaussian, random field. As such it actively contributes to the total energy density perturbations as well as to the anisotropic stress perturbation. Furthermore the Lorentz term changes the baryon velocity. 
This has important implications for the linear matter power spectrum which will be the focus of this work.
The linear matter power spectrum provides the initial distribution of the density field from which nonlinear structure evolves. 
It determines implicitly the distribution of neutral hydrogen in the post recombination universe during the cosmic dark ages
and later on ionized hydrogen.
Cosmic dawn starts with the formation of the first star forming galaxies within dark matter halos. These are sources of high energetic UV  photons and at later epochs X-ray photons from quasars that ionize and heat matter. These high energetic photons can redshift down to the corresponding Lyman $\alpha$ wave length
which can be absorbed by an hydrogen atom and emitted spontaneously allowing for the atom to change from, say, the hyperfine singlet to the triplet state which is the Wouthuysen-Field mechanism coupling the spin and gas temperature (e.g. \cite{Pritchard:2011xb}). At some point the Ly$\alpha$ coupling saturates, by which time the gas has heated above the temperature of the cosmic microwave background (CMB). The 21 cm line signal is the change in the brightness temperature of the CMB as seen by an observer today. It is necessary for a non zero signal that the spin temperature which determines the equilibrium of the ratio of the occupation numbers in the ground state hyperfine states of neutral hydrogen and the CMB temperature are different.
At large redshifts the gas is still cold. Thus the 21 cm line signal is seen in absorption. Once the heating of the gas  due to the high energetic photons in the UV and X-ray range becomes efficient the matter temperature is well above the CMB temperature and the 21 cm line signal is seen in emission. Moreover, in this case the change in brightness temperature will saturate.
Magnetic fields can also influence the 21cm line signal by additional heating of matter because of dissipative processes (cf. \cite{Tashiro:2006uv,Schleicher:2008hc,Sethi:2009dd}). However, here the focus will be on the effects due to the change in the linear matter power spectrum.

The magnetic field is assumed to be a non helical, gaussian random field determined by its two point function in $k$-space,
\begin{eqnarray}
\langle B_i^*(\vec{k})B_j(\vec{q})\rangle=(2\pi)^3\delta({\vec{k}-\vec{q}})P_B(k)\left(\delta_{ij}-\frac{k_ik_j}{k^2}\right),
\end{eqnarray}
where the power spectrum, $P_B(k)$ is given by \cite{kk1}
\begin{eqnarray}
P_B(k,k_m,k_L)=A_B\left(\frac{k}{k_L}\right)^{n_B}W(k,k_m)
\end{eqnarray}
where $k_L$ is a pivot wave number chosen to be 1 Mpc$^{-1}$ and  $W(k,k_m)=\pi^{-3/2}k_m^{-3}e^{-(k/k_m)^2}$
is a gaussian window function. $k_m$ corresponds to the largest scale
damped due to radiative viscosity before decoupling \cite{sb,jko}. $k_m$ has its largest value at recombination 
\begin{eqnarray}
k_m=299.66\left(\frac{B}{\rm nG}\right)^{-1}{\rm Mpc}^{-1}
\end{eqnarray}
for the bestfit parameters of Planck13+WP data  \cite{kuko15,Planck_cosmology}.

\section{The linear matter power spectrum}
\label{s1}
\setcounter{equation}{0}

At the epochs of interest here close to reionization the universe is matter dominated. 
The initial linear matter power spectrum is assumed to be given by  the contributions from the primordial curvature mode as well as the magnetic mode.
For modes inside the 
horizon the linear matter power spectrum of the adiabatic curvature mode is given by (cf., e.g.,\cite{hu,hw1,kk-secAniso})
\begin{eqnarray}
P^{(ad)}_m(k)=\frac{2\pi^2}{k^3}\left(\frac{k}{a_0H_0}\right)^4\frac{4}{25}A_s\left(\frac{k}{k_p}\right)^{n_s-1}T^2(k),
\end{eqnarray}
where the transfer function $T(k)$ is given by \cite{pu,bbks}
\begin{eqnarray}
T(k)=\frac{\ln(1+2.34q)}{2.34q}\left[1+3.89q+(16.1q)^2+(5.46q)^3+(6.71q)^4\right]^{-\frac{1}{4}}
\end{eqnarray}
where $q=\frac{k}{\Omega_{m,0}h^2 {\rm Mpc}^{-1}}$.

For the magnetic mode the matter power spectrum is found to be \cite{kk-secAniso}
\begin{eqnarray}
P_m^{(B)}(k)=\frac{2\pi^2}{k^3}\left(\frac{k}{a_0H_0}\right)^4\frac{4}{225}(1+z_{dec})^2\left(\frac{\Omega_{\gamma,0}}{\Omega_{m,0}}\right)^2
{\cal P}_L(k),
\end{eqnarray}
where ${\cal P}_L(k)$ is the dimensionless power spectrum determining the two point function of the Lorentz term
$\langle L^*(\bm{k})L(\bm{k'})\rangle=\frac{2\pi^2}{k^3}{\cal P}_L(k)$ given by \cite{kk1}
\begin{eqnarray}
{\cal P}_L(k)&=&\frac{9}{\left[\Gamma\left(\frac{n_B+3}{2}\right)\right]^2}\left(\frac{\rho_{B,0}}{\rho_{\gamma,0}}\right)^2
\left(\frac{k}{k_m}\right)^{2(n_B+3)}e^{-\left(\frac{k}{k_m}\right)^2}\nonumber\\
&\times&\int_0^{\infty}dz z^{n_B+2}e^{-2\left(\frac{k}{k_m}\right)^2z^2}\int_{-1}^1 dxe^{2\left(\frac{k}{k_m}\right)^2zx}
(1-2zx+z^2)^{\frac{n_B-2}{2}}\nonumber\\
&\times&\left[1+2z^2+(1-4z^2)x^2-4zx^3+4z^2x^4\right],
\label{pL}
\end{eqnarray}
and $x\equiv\frac{\bm{k}\cdot\bm{q}}{kq}$ and $z\equiv\frac{q}{k}$ where $\bm{q}$ is the wave number over which the 
resulting convolution integral is calculated.

\begin{figure}[h!]
\centerline{\epsfxsize=3.0in\epsfbox{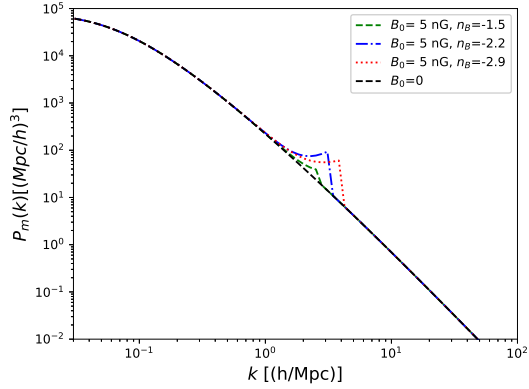}}
\caption{Total linear matter power spectrum at the present epoch for different values of the magnetic field parameters and the best fit Planck13+WP parameters \cite{Planck_cosmology}.}
\label{fig1}
\end{figure}

The resulting linear matter power  spectrum for the magnetic plus adiabatic mode is shown in figure \ref{fig1}.
Below the magnetic Jeans scale pressure supports against collapse and prevents any further growth of the density perturbation. Therefore the 
linear matter power spectrum of the magnetic mode is cut-off at the wave number corresponding to the magnetic Jeans scale $k_J$, \cite{sesu}
\begin{eqnarray}
\left(\frac{k_J}{{\rm Mpc}^{-1}}\right)=\left[14.8\left(\frac{\Omega_m}{0.3}\right)^{\frac{1}{2}}\left(\frac{h}{0.7}\right)
\left(\frac{B}{10^{-9}{\rm G}}\right)^{-1}\left(\frac{k_L}{{\rm Mpc}^{-1}}\right)^{\frac{n_B+3}{2}}\right]^{\frac{2}{n_B+5}}.
\end{eqnarray}
The linear matter power spectrum is normalized to $\sigma_8$ of the best fit Planck13+WP parameters \cite{Planck_cosmology}.

\section{The 21 cm line signal}
\label{s2}
\setcounter{equation}{0}

For the simulations the {\tt Simfast21}\footnote{https://github.com/mariogrs/Simfast21}
code \cite{simfast21-1,simfast21-2} is adapted to allow for reading in the modified linear matter power spectra.
{\tt Simfast21} calculates the change in the brightness temperature following a similar algorithm as the {\tt 21cmFAST}
\footnote{http://homepage.sns.it/mesinger/DexM\underline{\hspace{0.2cm}}21cmFAST.html}
 code \cite{21cmfast}.
The initial Gaussian, random density field is determined by the linear matter power spectrum. 
The subsequent evolution in time leads to gravitational collapse and nonlinear structure and formation of dark matter halos.
The halo distribution is found by using the excursion formalism whereby  a given region is considered to undergo gravitational collapse if its mean overdensity is larger than a certain critical value $\delta_c(M,z)$ depending on the halo mass $M$ and redshift $z$.
As the halo positions are based on the linear density field these have to be corrected for the effects of the non linear dynamics. This is done using the Zel'dovich approximation.
A source for reionization of matter in the universe are galaxies which form inside dark matter haloes. 
Thus the corrected halo distribution allows to determine the ionization regions.
In the version of {\tt Simfast21} \cite{simfast21-2} used in this work the criterion to decide whether a given region is ionized is determined by the local ionization rate $R_{ion}$ and the recombination rate $R_{rec}$. These are implemented using a numerical fitting formula which was obtained
from numerical simulations. 
In addition there is a free paramter which is the assumed escape fraction of ionizing photons from star forming regions $f_{esc}$.
A bubble cell is defined to be completely ionized if the condition 
\begin{eqnarray}
f_{esc}R_{ion}\geq R_{rec}
\label{ion}
\end{eqnarray}
 is satisfied. 
In this work the value of the escape rate is set to $f_{esc}=0.06$.
Once the evolution of the ionization field has been determined the 21cm line signal can be calculated.
In equilibrium the ratio of the  populations of the two hyperfine states, the less energetic singlet state and the more energetic triplet state,  of neutral hydrogen is determined by the spin temperature $T_S$, e.g. \cite{LF, MBW},
\begin{eqnarray}
\frac{n_1}{n_0}=\left(\frac{g_1}{g_0}\right)\exp\left(-\frac{T_*}{T_S}\right),
\end{eqnarray}
where $T_*=E_{10}/k_B=68$ mK and the energy difference $E_{10}$ corresponds to a wave length $\lambda_{10}\sim 21$ cm.
When CMB photons travel through a medium with neutral hydrogen some of them will be absorbed by hydrogen atoms in the singlet state exciting them to the triplet state. At the same time hydrogen in the triplet state might spontaneously relax to the singlet state emitting a photon.
Therefore the observed brightness temperature of the CMB results in 
\begin{eqnarray}
T_b(z)=T_{CMB}(z)e^{-\tau(z)}+\left(1-e^{-\tau(z)}\right)T_S(z)
\end{eqnarray}
where $T_{CMB}(z)$ is the brightness temperature of the CMB without absorption and
$\tau(z)$ is the corresponding optical depth along the ray through the medium. Thus the change in the brightness temperature of the CMB as measured by an observer today is given by, e.g. \cite{LF, MBW},
\begin{eqnarray}
\delta T_b=T_b-T_{CMB}=\frac{\left(1-e^{-\tau(z)}\right)\left[T_S(z)-T_{CMB}(z)\right]}{1+z}.
\end{eqnarray}
With the approximations $\tau\ll 1$ and $z\gg 1$, the 21cm line signal is given by
\begin{eqnarray}
\delta T_b= 28 {\rm mK}\left(\frac{\Omega_{b,0}h}{0.03}\right)\left(\frac{\Omega_{m,0}}{0.3}\right)^{-\frac{1}{2}}\left(\frac{1+z}{10}\right)^{\frac{1}{2}}
\frac{\left(T_S-T_{CMB}\right)}{T_S}x_{HI}.
\label{dTb}
\end{eqnarray}
As can be seen from equation (\ref{dTb}) there is only a signal if the spin temperature is different from the CMB radiation temperature.
Otherwise the hydrogen spin state is in thermal equilibrium with the CMB and emission and absorption processes are compensated on average.
The net emission or absorption result from a higher or lower, respectively, spin temperature than the CMB radiation temperature.
There are several processes which can lead to the spin temperature being different from the CMB temperature such as the presence of radiation sources or heating of the gas. In addition there are two processes which can change the spin temperature of the neutral hydrogen gas. Firstly, collisional excitation and de-excitation of the spin states. Secondly the Wouthuysen-Field process which couples the two spin states.  
In the limit that the spin temperature $T_S$ is much higher than the temperature of the CMB photons $T_{CMB}$ the change in the brightness temperature $\delta T_b$ (cf. equation (\ref{dTb})) becomes saturated.
This is the case for lower redshifts when UV photons from star forming galaxies heat the IGM.
For simplicity we will assume here that $T_{S}\gg T_{CMB}$.
The focus here is to study the effect of the presence of a primordial magnetic field on the 21 cm line signal induced by the change in the linear matter power spectrum.

\section{Results}
\label{s3}
\setcounter{equation}{0}

\begin{figure}[ht]
\centerline{
\epsfxsize=2.3in\epsfbox{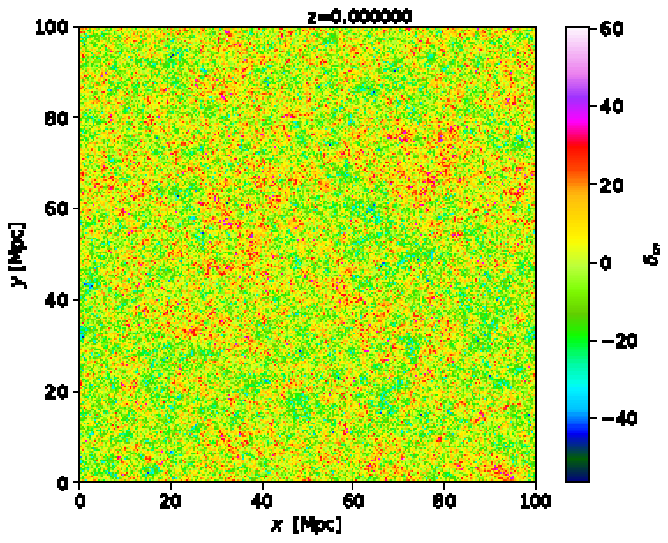}
%\hspace{0.1cm}
\epsfxsize=2.3in\epsfbox{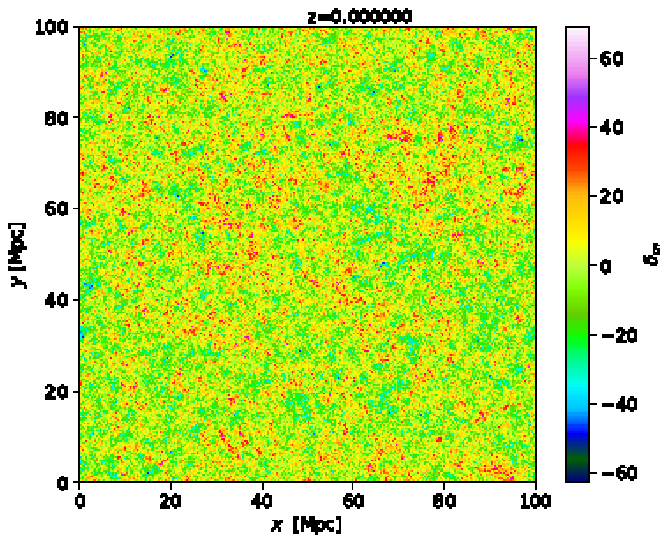}
%\hspace{0.1cm}
\epsfxsize=2.3in\epsfbox{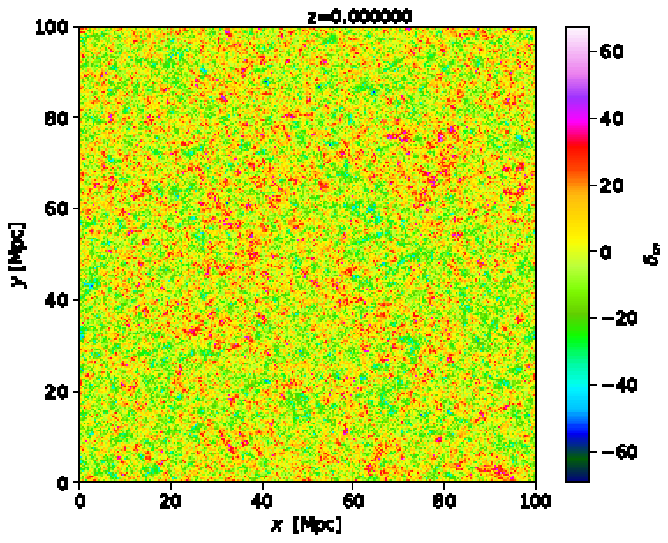}}
\centerline{
\epsfxsize=2.3in\epsfbox{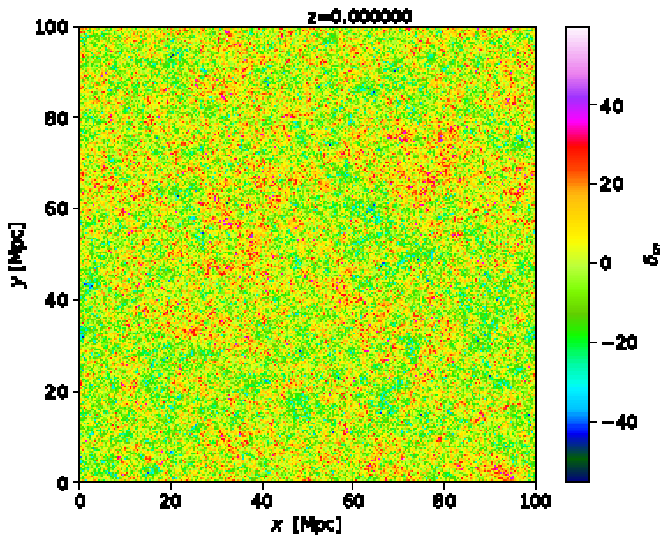}
%\hspace{0.1cm}
\epsfxsize=2.3in\epsfbox{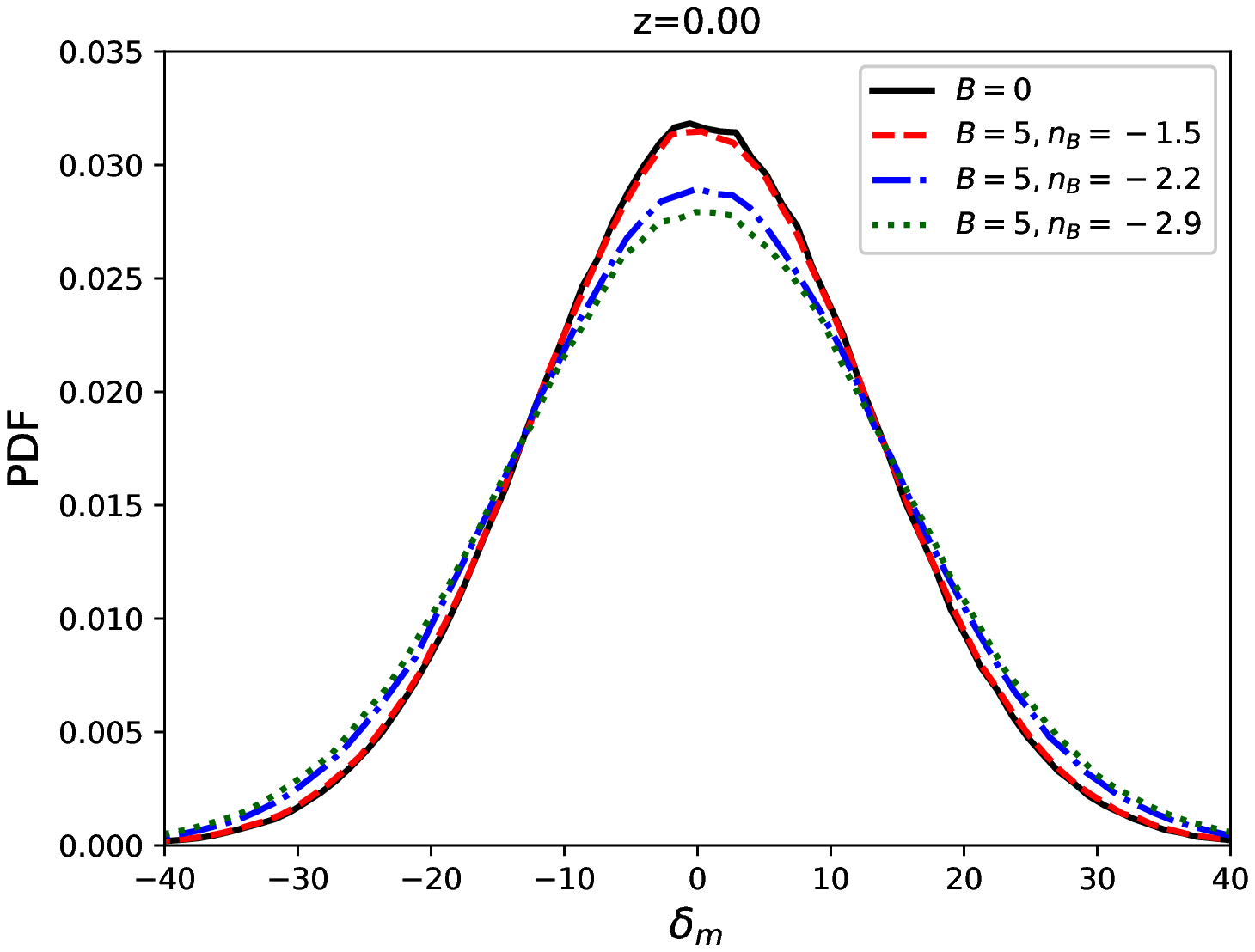}
}
\caption{The density field at $z=0$ when linear evolution upto the present is assumed. The upper panel shows from {\sl left} to {\sl right}
the simulations for $B_0=5$ nG and the spectral magnetic indices $n_B=-1.5$, $n_B=-2.2$ and $n_B=-2.9$. The lower panel shows the simulation of a pure adiabatic mode ({\sl left}) and the PDF for all four cases ({\sl right}).} 
\label{fig2}
\end{figure}

\begin{figure}[ht]
\centerline{
\epsfxsize=2.0in\epsfbox{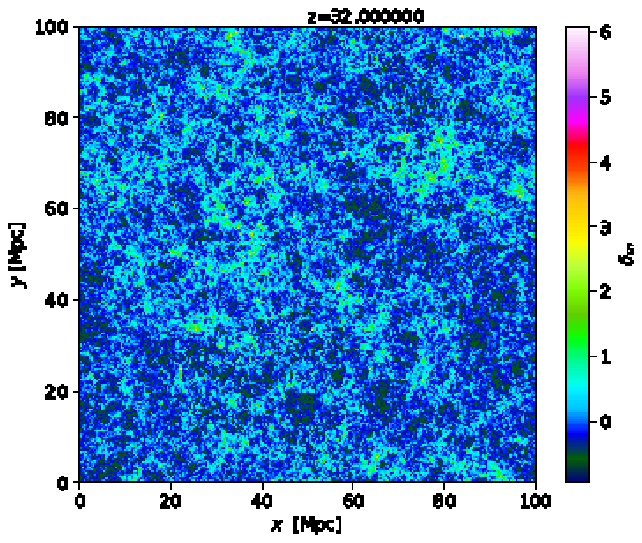}
\epsfxsize=2.0in\epsfbox{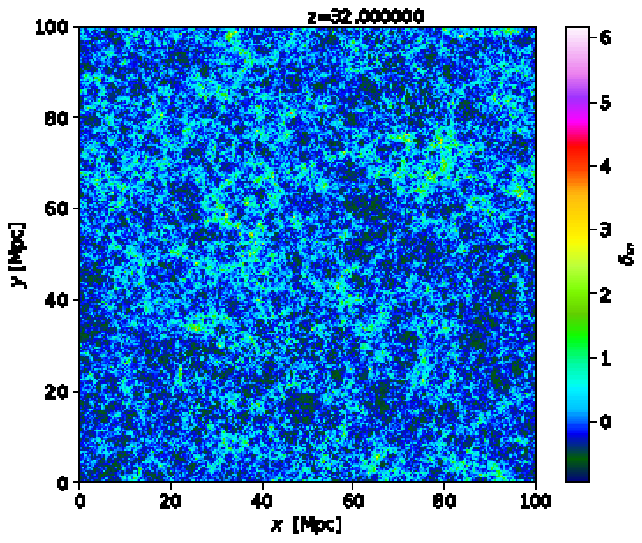}
%\hspace{0.1cm}
\epsfxsize=2.in\epsfbox{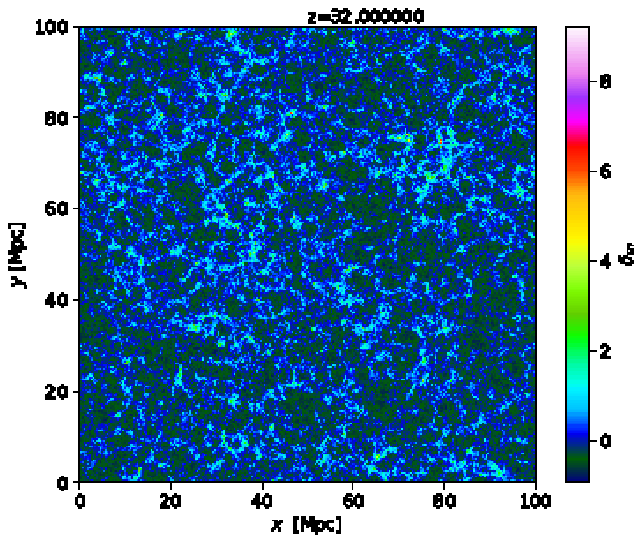}
%\hspace{0.1cm}
\epsfxsize=2.in\epsfbox{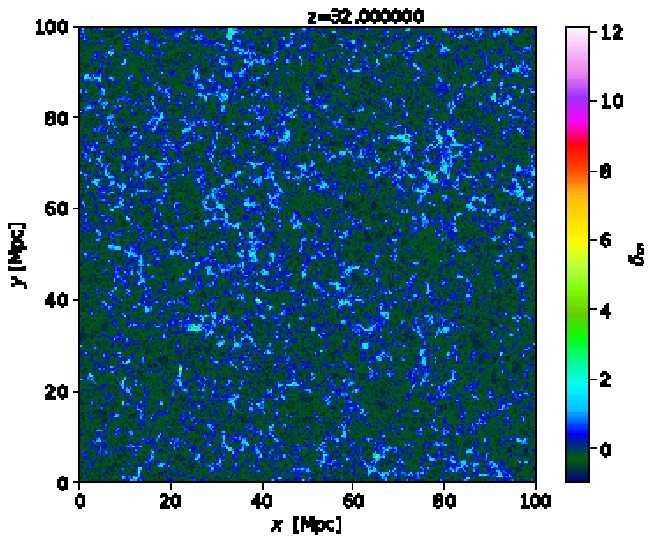}}
\centerline{
\epsfxsize=2.0in\epsfbox{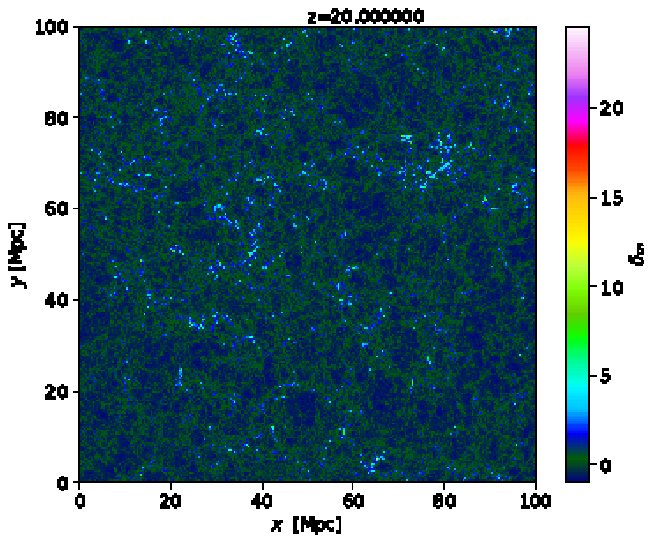}
\epsfxsize=2.0in\epsfbox{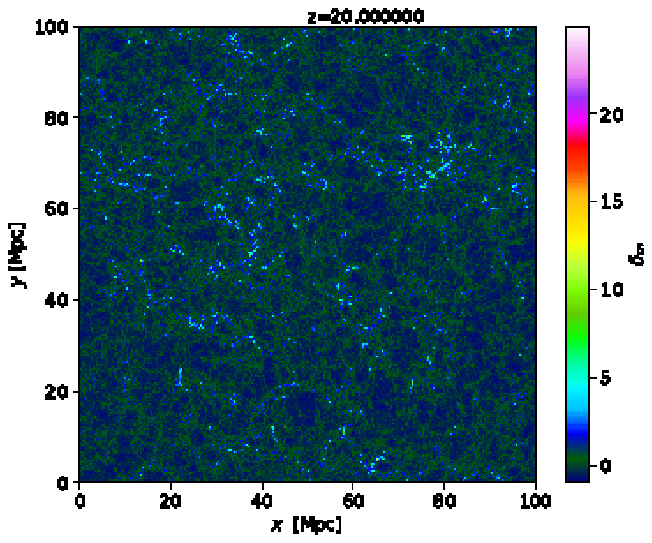}
%\hspace{0.1cm}
\epsfxsize=2.in\epsfbox{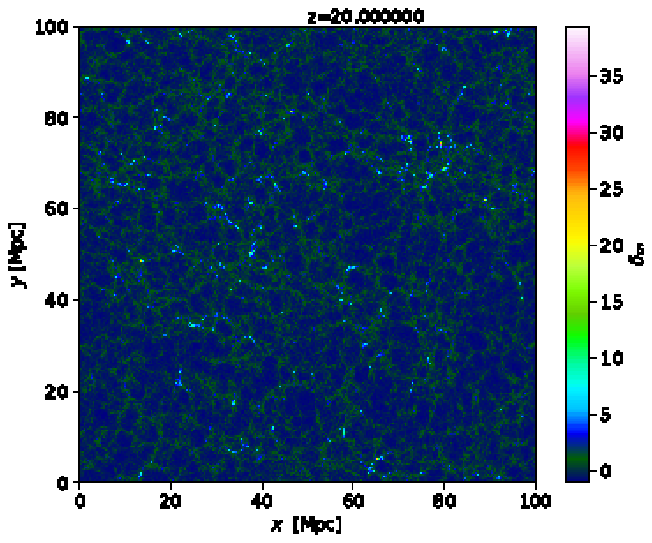}
%\hspace{0.1cm}
\epsfxsize=2.in\epsfbox{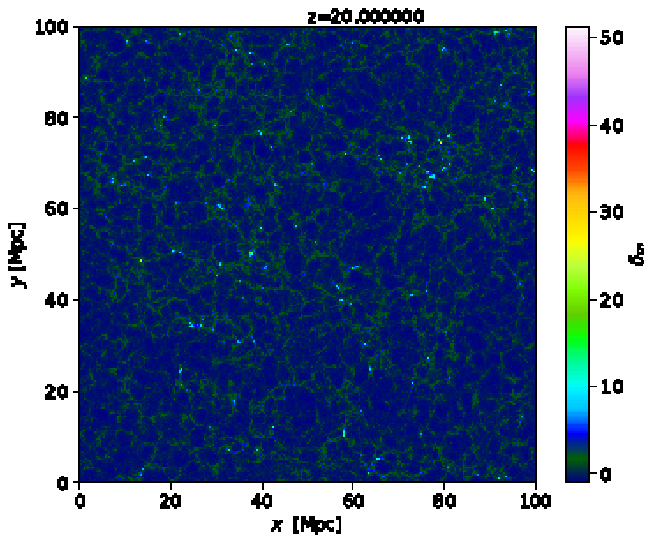}
}
\centerline{
\epsfxsize=2.0in\epsfbox{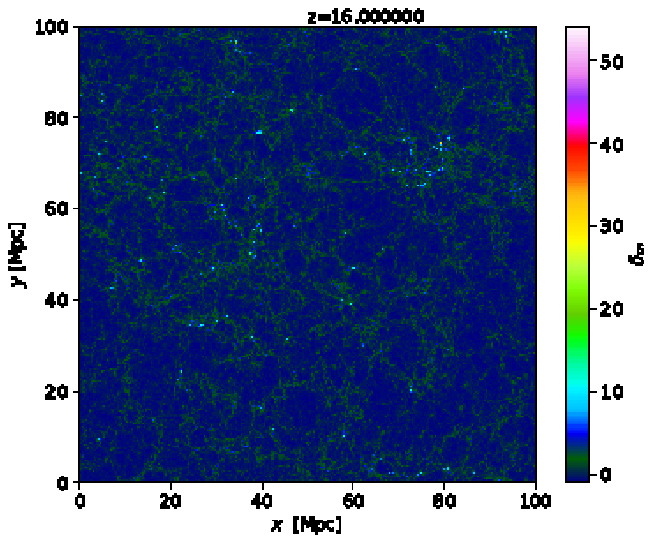}
\epsfxsize=2.0in\epsfbox{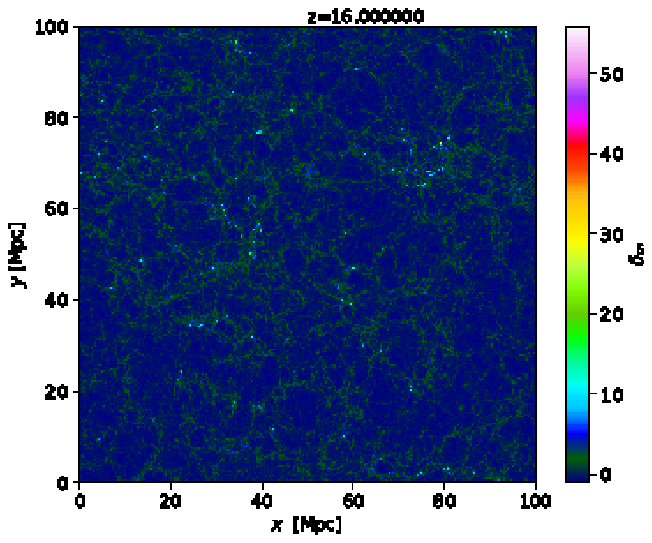}
%\hspace{0.1cm}
\epsfxsize=2.in\epsfbox{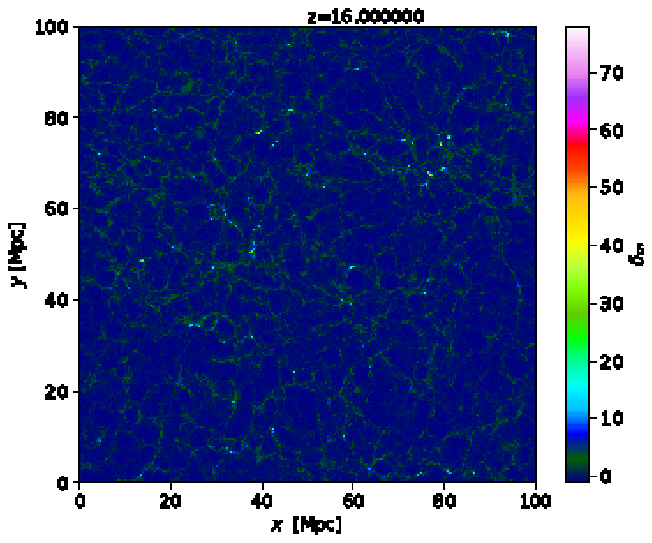}
%\hspace{0.1cm}
\epsfxsize=2.in\epsfbox{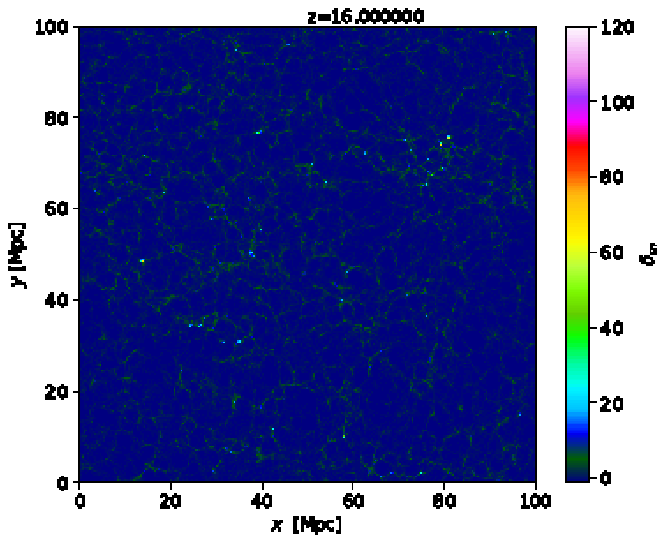}
}
\centerline{
\epsfxsize=2.0in\epsfbox{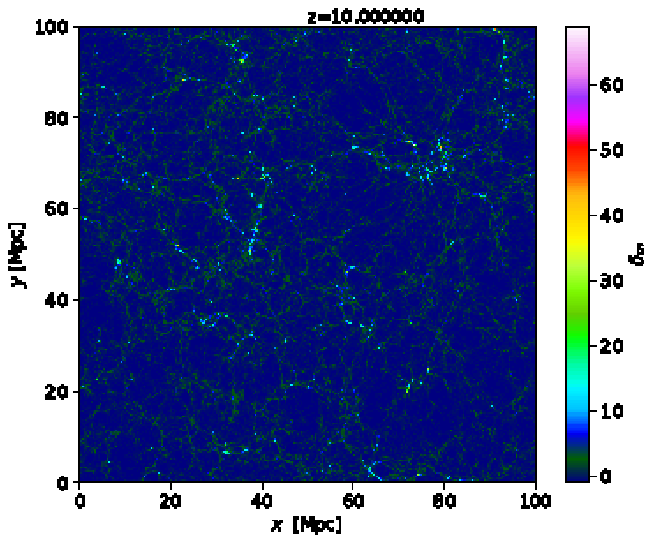}
\epsfxsize=2.0in\epsfbox{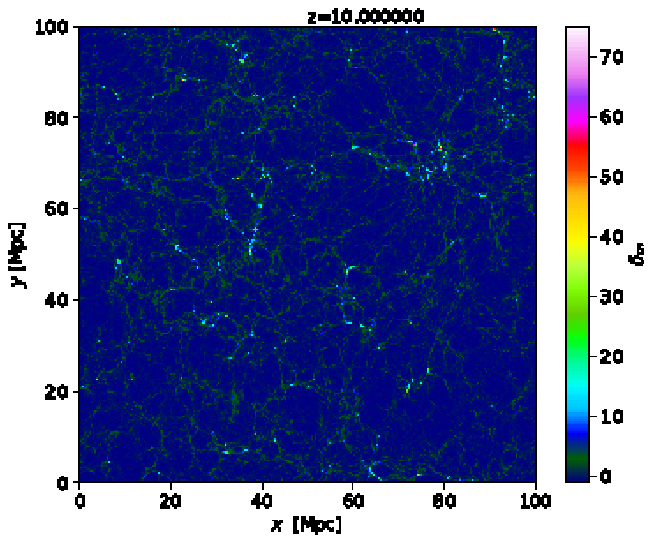}
%\hspace{0.1cm}
\epsfxsize=2.in\epsfbox{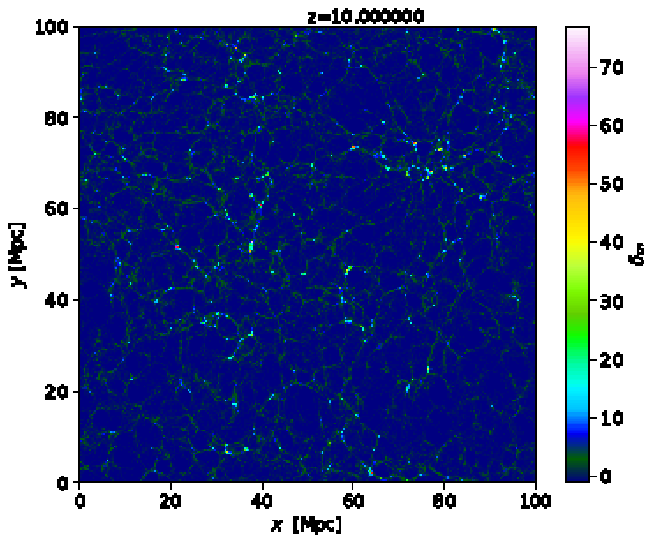}
%\hspace{0.1cm}
\epsfxsize=2.in\epsfbox{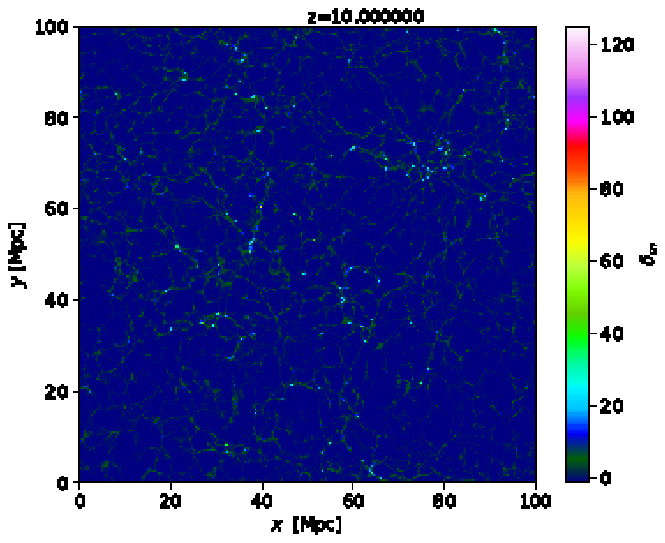}
}

\caption{The density field at $z=32,20,16,10$ with the highest redshift shown in the first row and the lowest redshift in the last row. In each panel from {\sl left} to {\sl right} are shown
the simulations for $B_0=0$ ({\sl far left}) and subsequently for the cases
$B_0=5$ nG and the spectral magnetic indices $n_B=-1.5$, $n_B=-2.2$ and $n_B=-2.9$.
} 
\label{fig3}
\end{figure}

\begin{figure}[ht]
\centerline{
\epsfxsize=2.0in\epsfbox{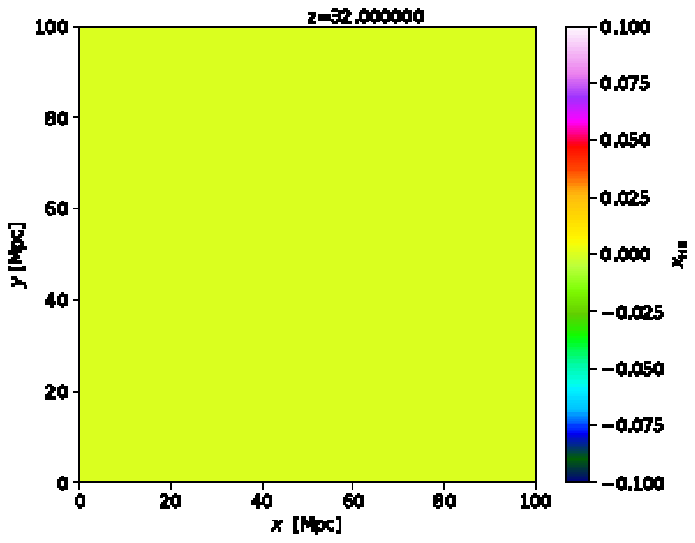}
\epsfxsize=2.0in\epsfbox{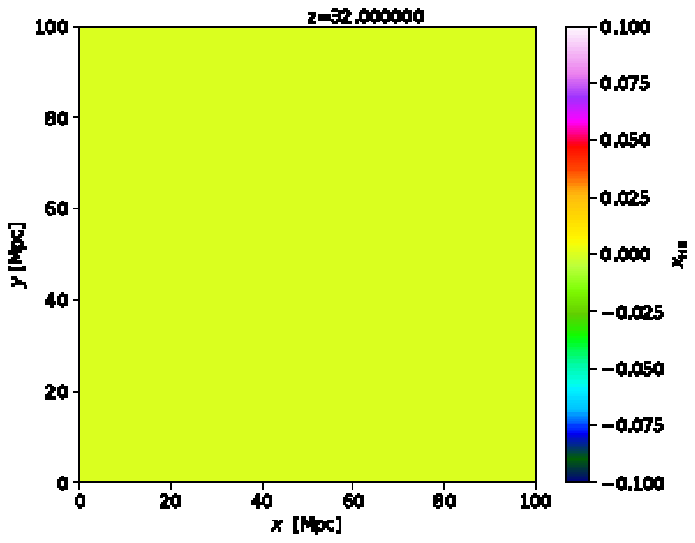}
%\hspace{0.1cm}
\epsfxsize=2.in\epsfbox{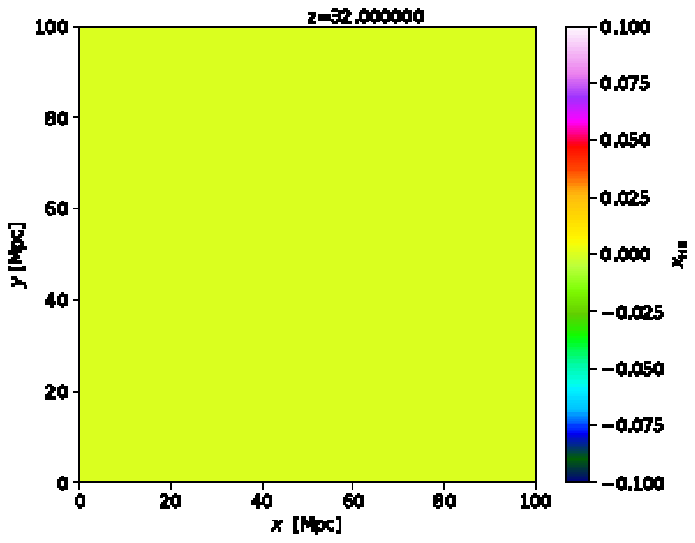}
%\hspace{0.1cm}
\epsfxsize=1.95in\epsfbox{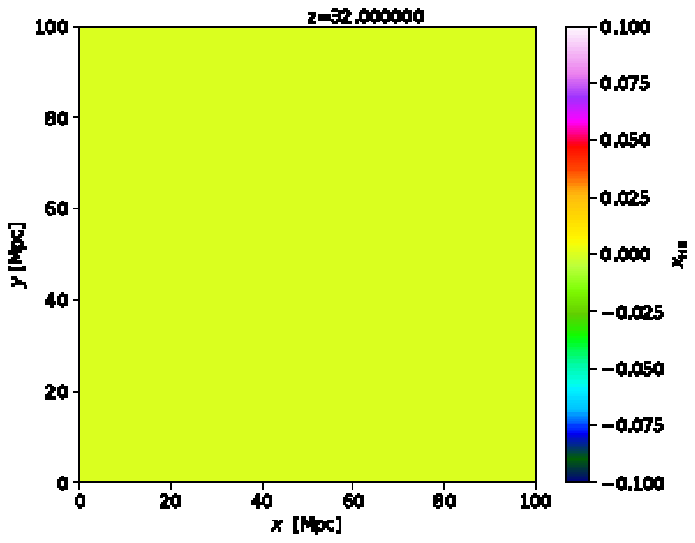}}
\centerline{
\epsfxsize=2.0in\epsfbox{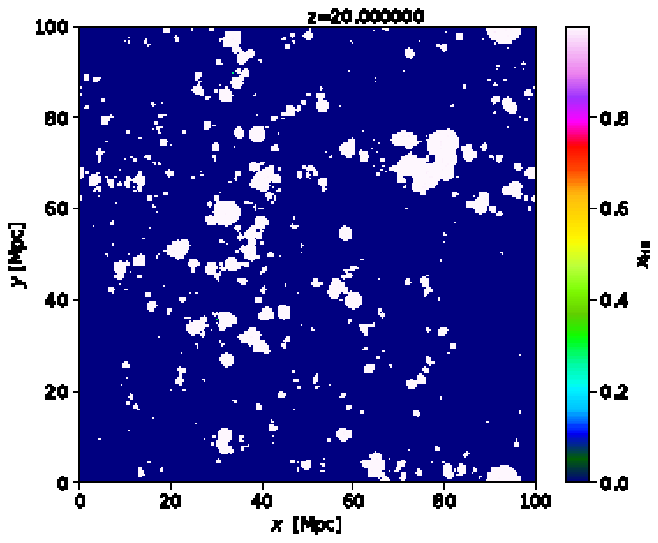}
\epsfxsize=2.0in\epsfbox{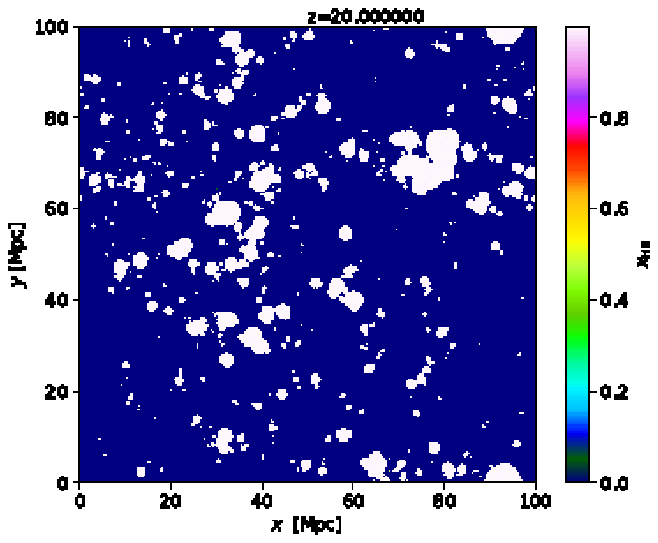}
%\hspace{0.1cm}
\epsfxsize=2.in\epsfbox{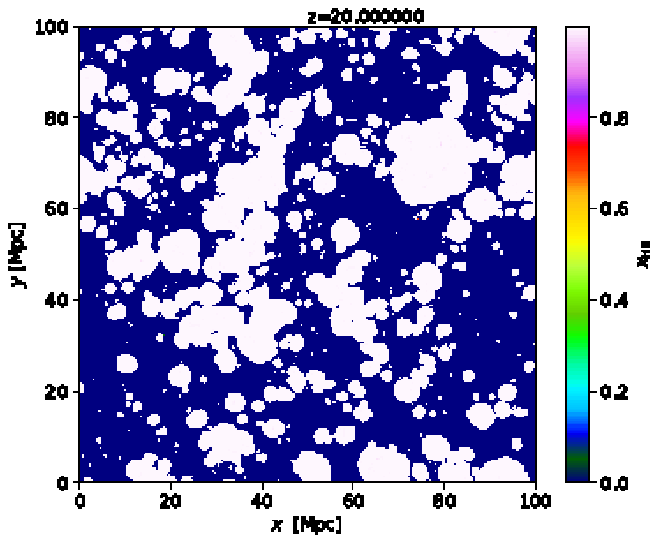}
%\hspace{0.1cm}
\epsfxsize=2.in\epsfbox{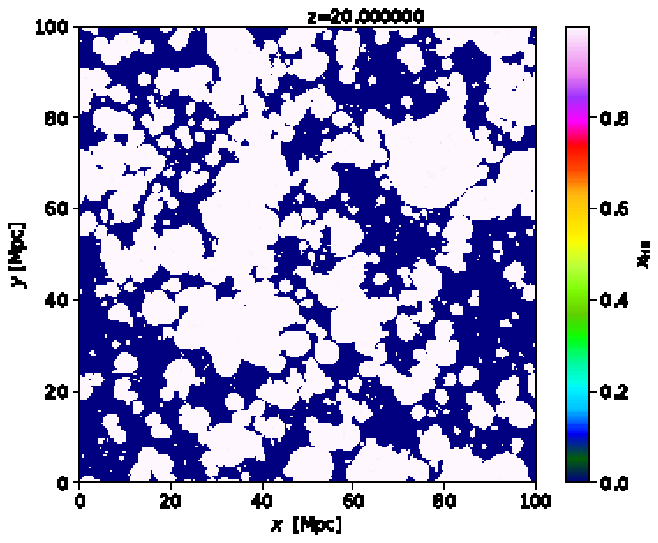}
}
\centerline{
\epsfxsize=2.0in\epsfbox{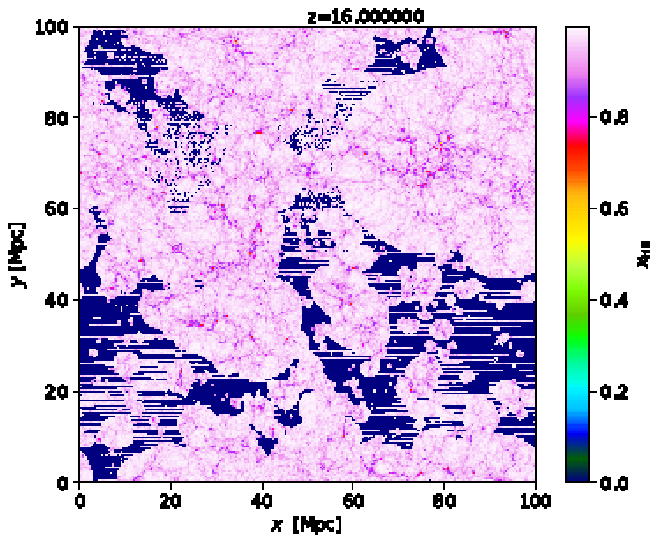}
\epsfxsize=2.0in\epsfbox{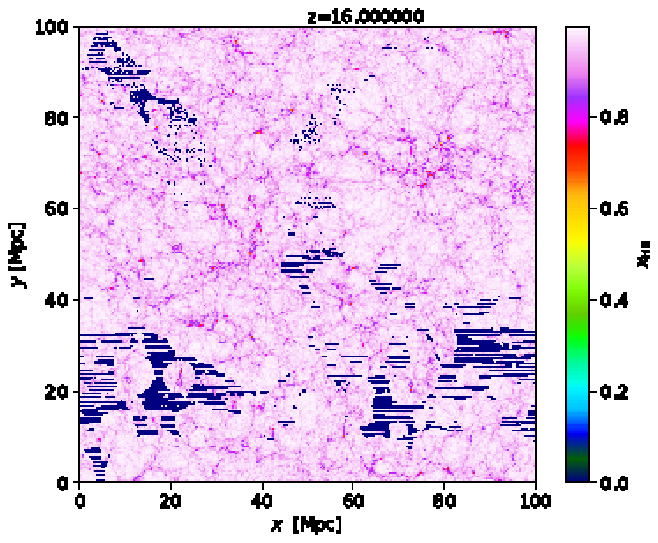}
%\hspace{0.1cm}
\epsfxsize=2.in\epsfbox{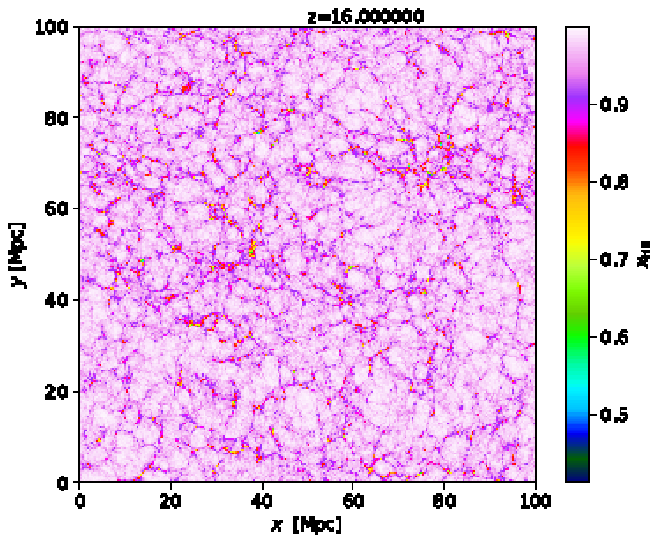}
%\hspace{0.1cm}
\epsfxsize=2.in\epsfbox{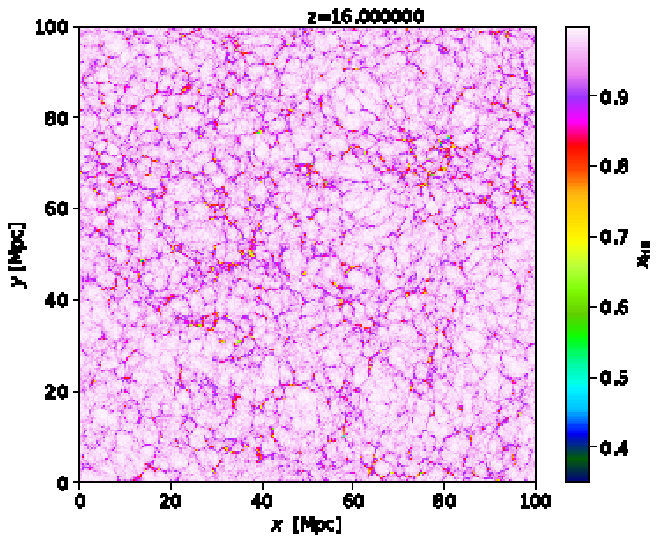}
}
\centerline{
\epsfxsize=2.0in\epsfbox{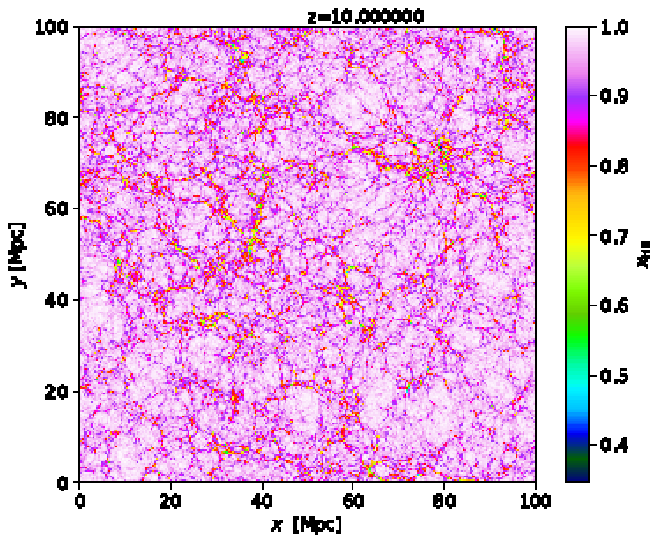}
\epsfxsize=2.0in\epsfbox{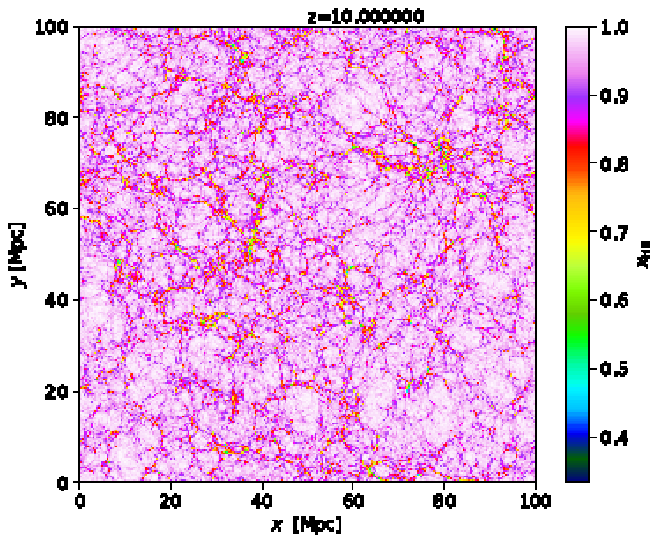}
%\hspace{0.1cm}
\epsfxsize=2.in\epsfbox{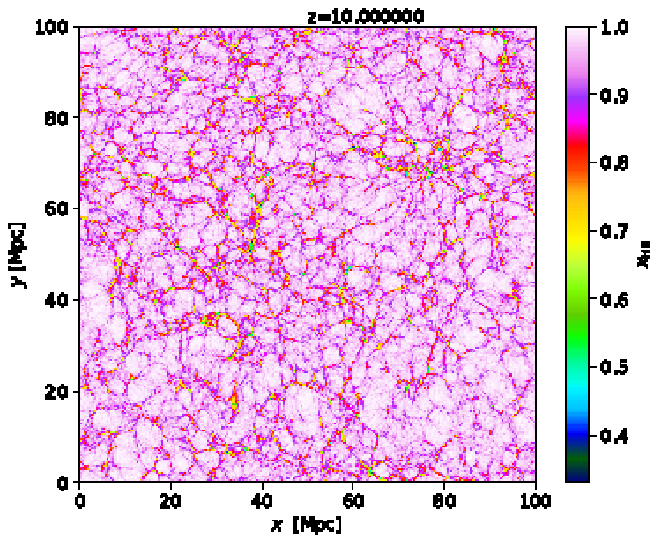}
%\hspace{0.1cm}
\epsfxsize=2.in\epsfbox{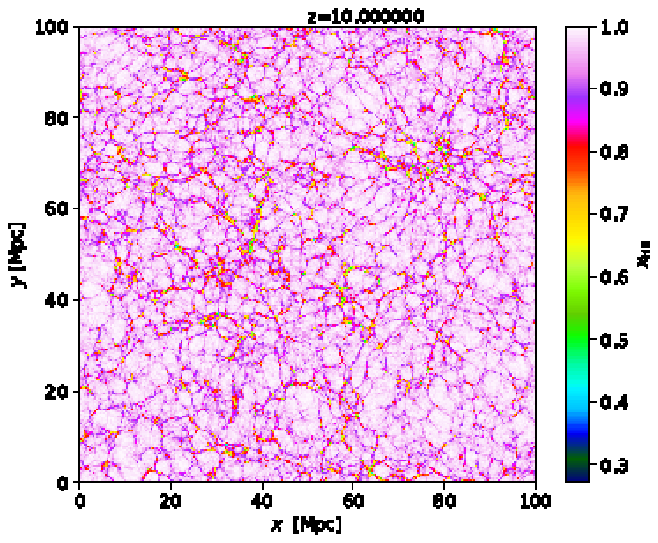}
}
\caption{HII regions shown  at $z=32,20,16,10$ with the highest redshift shown in the first row and the lowest redshift in the last row. In each panel from {\sl left} to {\sl right} are shown
the simulations for $B_0=0$ ({\sl far left}) and subsequently for the cases
$B_0=5$ nG and the spectral magnetic indices $n_B=-1.5$, $n_B=-2.2$ and $n_B=-2.9$.
} 
\label{figxHII}
\end{figure}

\begin{figure}[ht]
\centerline{
\epsfxsize=3.0in\epsfbox{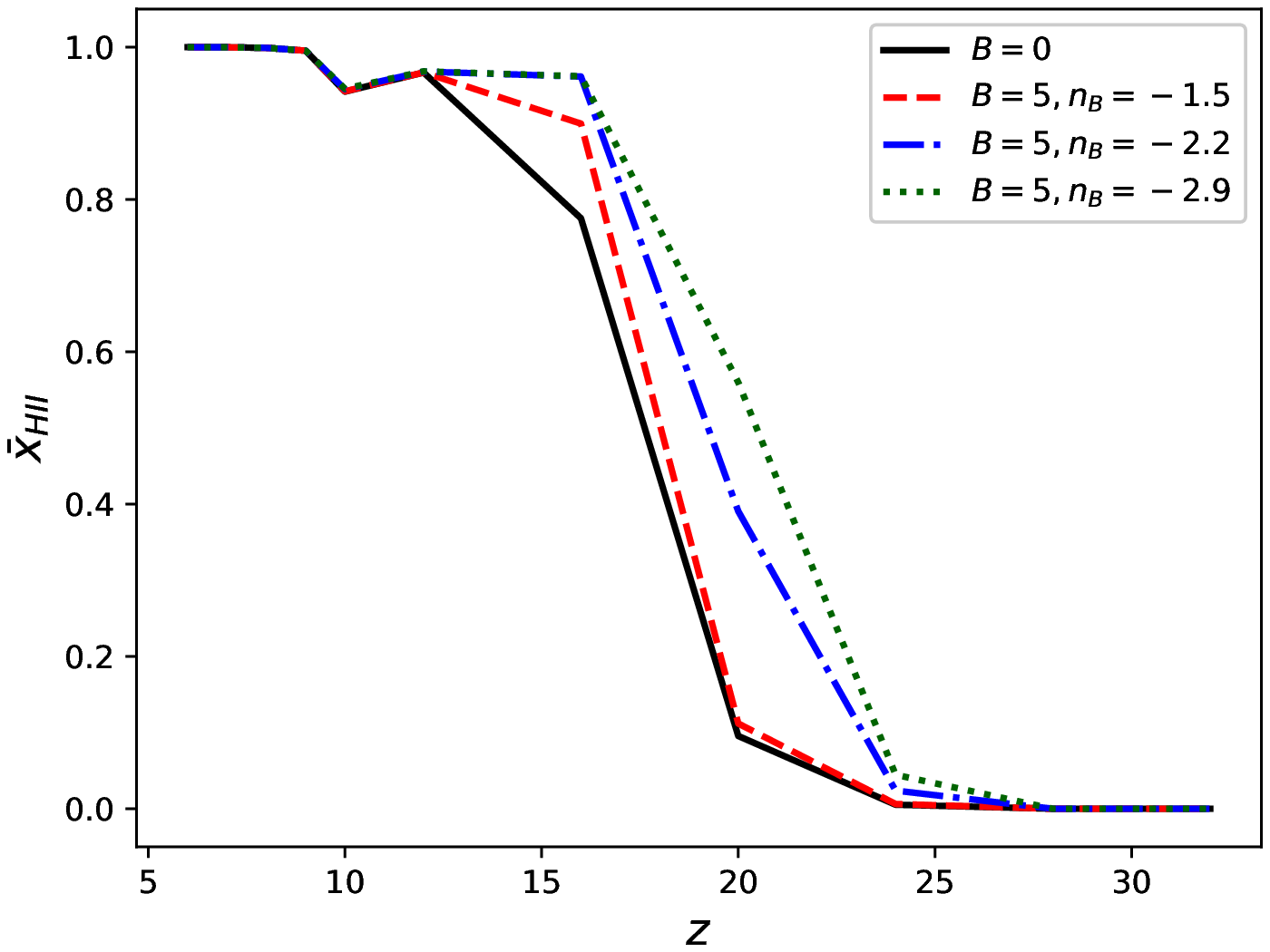}
}
\caption{The average ionization fraction as a function of redshift obtained from the simulation boxes.
} 
\label{figmeanxHII}
\end{figure}

In the numerical simulations the total linear matter power spectrum as calculated in section \ref{s1} is used.
In figure \ref{fig2} the density field at $z=0$ is shown when a linear evolution  is assumed.
The simulation boxes with each side corresponding to 100 Mpc are shown for no magnetic field, and in the presence of a magnetic field of field strength 5 nG and spectral indices $n_B=-1.5$, $n_B=-2.2$ and $n_B=-2.9$ are shown. The last panel on the second line of figure \ref{fig2} shows the 
 corresponding probability density function (pdf) for all four cases manifesting the initially assumed Gaussian distribution.
For the visualization the python code {\tt tocmfastpy} 
\footnote{J. Prichard, https://github.com/pritchardjr/tocmfastpy} 
has been adapted.
The {\tt Simfast21} code uses the Zeldovich approximation to obtain the nonlinear density field from which the halo distribution is obtained.
The nonlinear density fields are shown for redshifts $z=32$ to $z=10$ in figure \ref{fig3}. The effect of the feature in the initial linear matter power spectrum manifests itself by an increase in structure and amplitude in the matter density field. 
In figure \ref{figxHII} the distribution of the ionized hydrogen regions are shown at different redshifts with and without the primordial magnetic field. At the largest redshift shown, $z=32$, reionization has not started yet and there is no trace of ionized hydrogen ({\sl upper panel}). 
The epoch of reionization (EoR) starts before redshift $z=20$. Ionized gas forms bubbles of increasing size. This corresponds to the classic inside-out topology where the densest regions are ionized first which is the underlying assumption of the {\tt Simfast21} code. This can be nicely seen when comparing, for example, 
 the nonlinear matter density fields and the distributions of the  ionized regions at a redshift $z=20$ ({\sl second panel from above}  in figures \ref{fig3}
 and \ref{figxHII}). Moreover, the effect of the matter density field modified by the presence of the
magnetic field varies visibly with the different choices of the the magnetic field spectral index $n_B$.
In figure \ref{figmeanxHII} the average ionization fraction of each simulation box is shown as a function of redshift $z$. 
The beginning of EoR  varies with the parameters of the magnetic field, with the magnetic field with the smallest spectral index, $n_B=-2.9$, having the largest effect.
This is a simplified vision of the evolution of the ionization fraction as shown in the simulations of figure \ref{figxHII}.
As can be appreciated from figure \ref{figmeanxHII} reionization is completed at a redshift below $z=10$.
From figure \ref{figmeanxHII} it is also interesting to note that models including magnetic fields 
with the smallest spectral index, $n_B=-2.9$, show the longest duration of EoR.
It starts at larger values of redshift than in the other cases but it still reaches completion only below redshifts $z$ less than 10.

\begin{figure}[ht]
\centerline{
\epsfxsize=2.0in\epsfbox{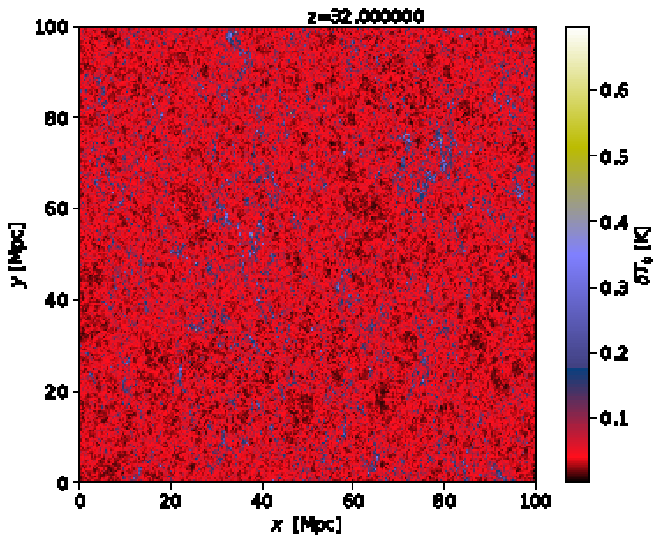}
\epsfxsize=2.0in\epsfbox{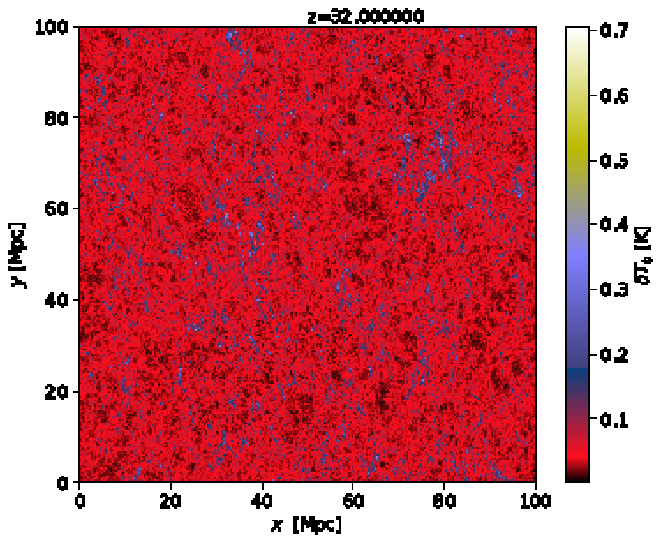}
%\hspace{0.1cm}
\epsfxsize=2.in\epsfbox{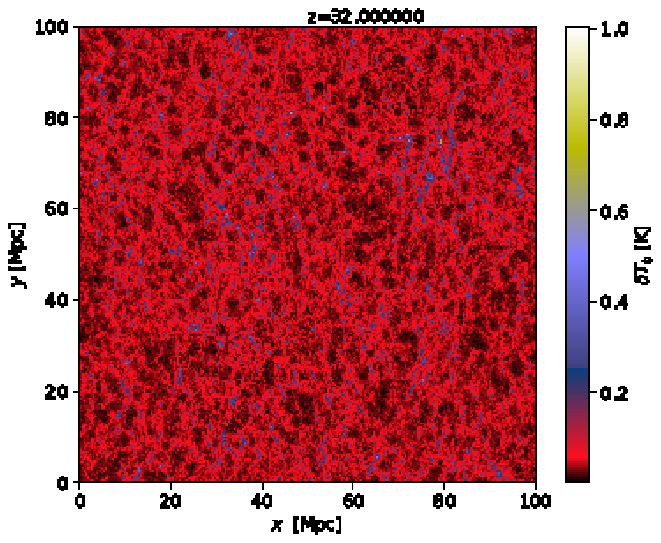}
%\hspace{0.1cm}
\epsfxsize=2.in\epsfbox{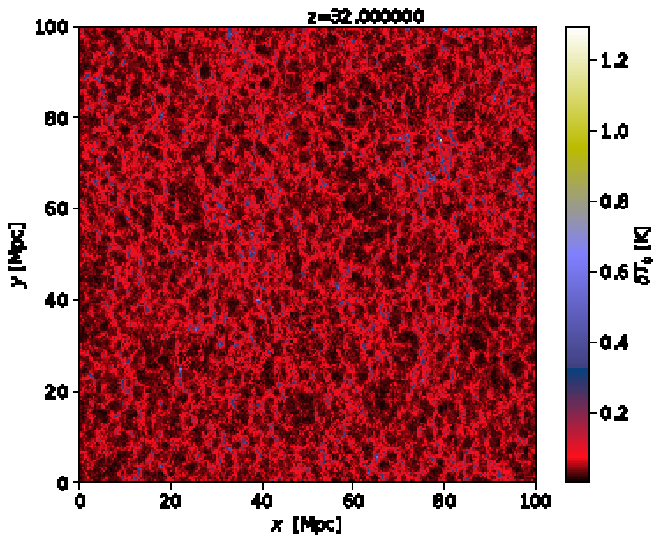}}
\centerline{
\epsfxsize=2.0in\epsfbox{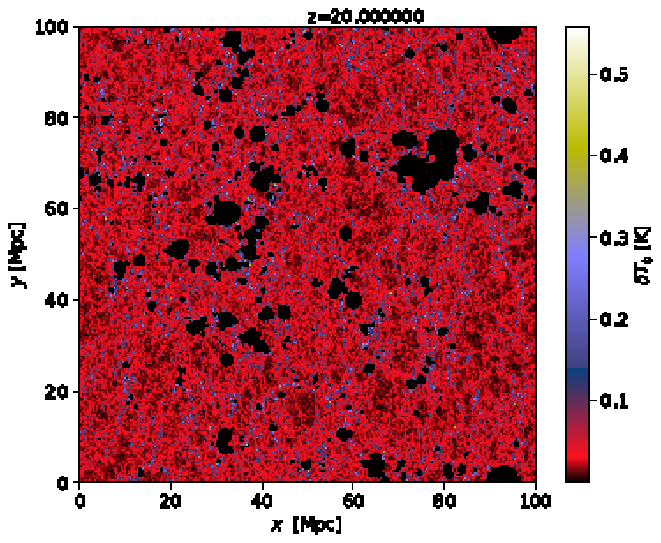}
\epsfxsize=2.0in\epsfbox{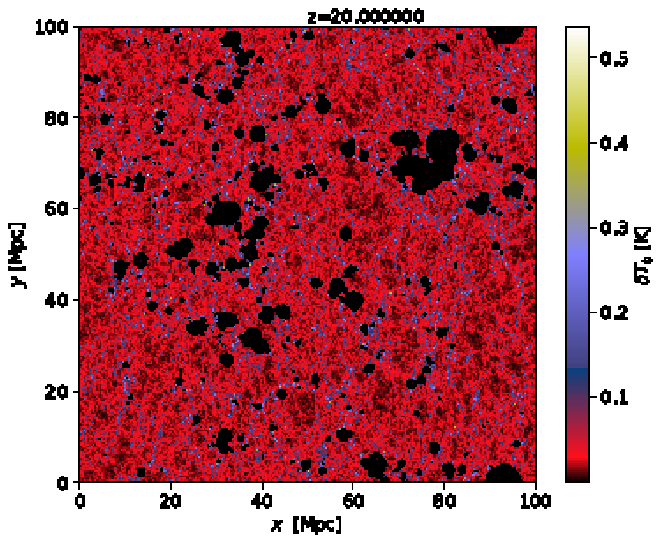}
%\hspace{0.1cm}
\epsfxsize=2.in\epsfbox{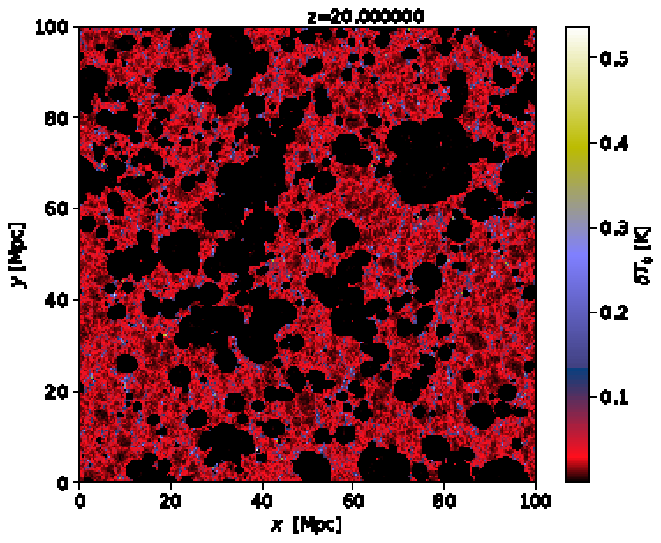}
%\hspace{0.1cm}
\epsfxsize=2.in\epsfbox{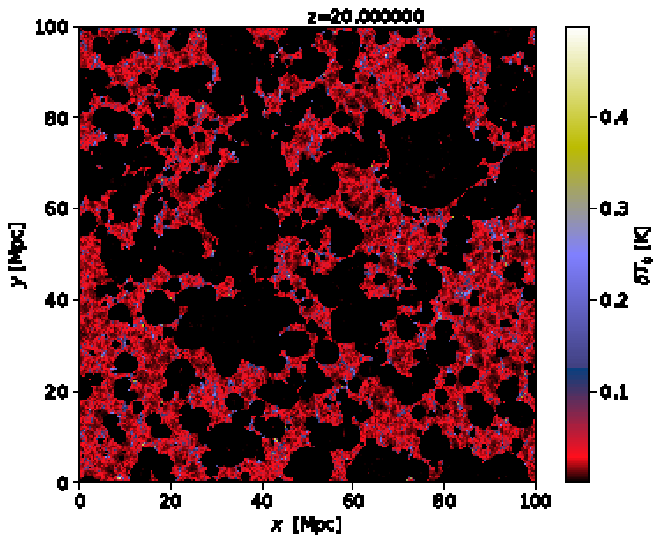}
}
\centerline{
\epsfxsize=2.0in\epsfbox{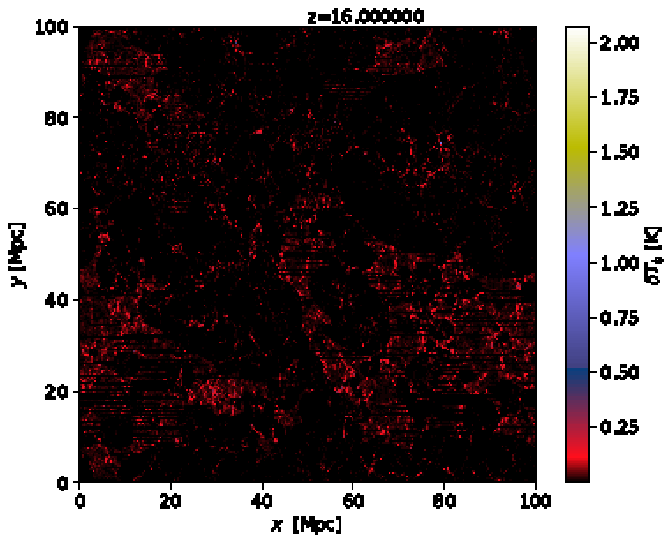}
\epsfxsize=2.0in\epsfbox{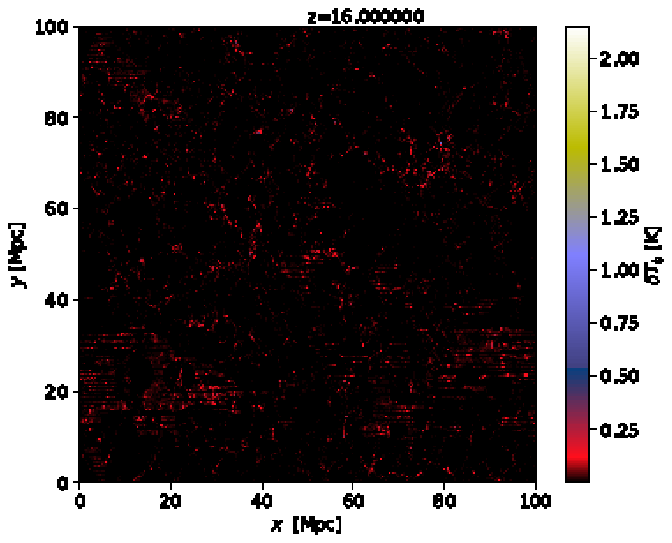}
%\hspace{0.1cm}
\epsfxsize=2.in\epsfbox{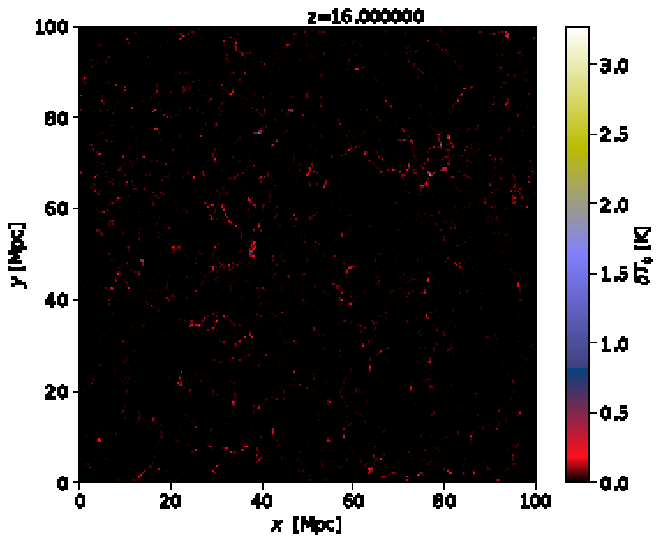}
%\hspace{0.1cm}
\epsfxsize=2.in\epsfbox{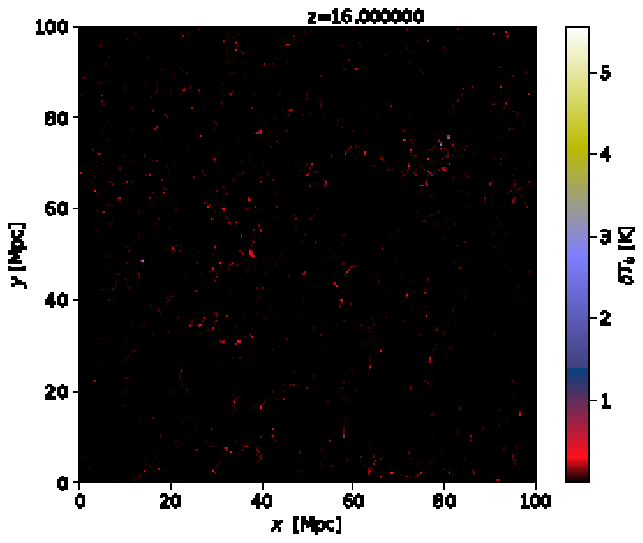}
}
\centerline{
\epsfxsize=2.0in\epsfbox{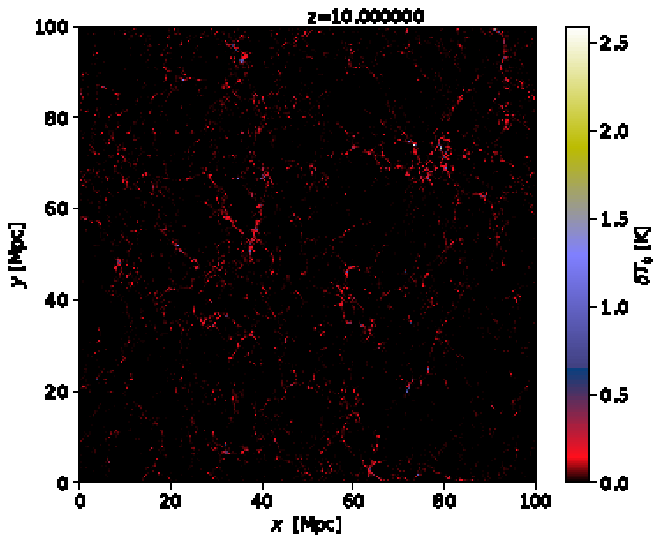}
\epsfxsize=2.0in\epsfbox{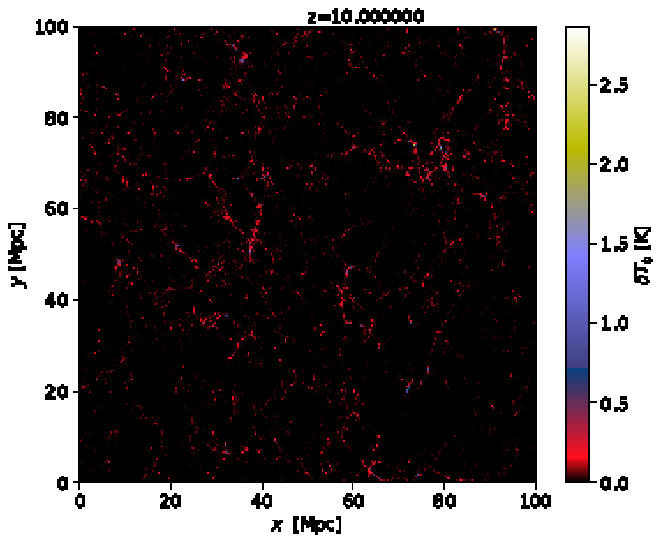}
%\hspace{0.1cm}
\epsfxsize=2.in\epsfbox{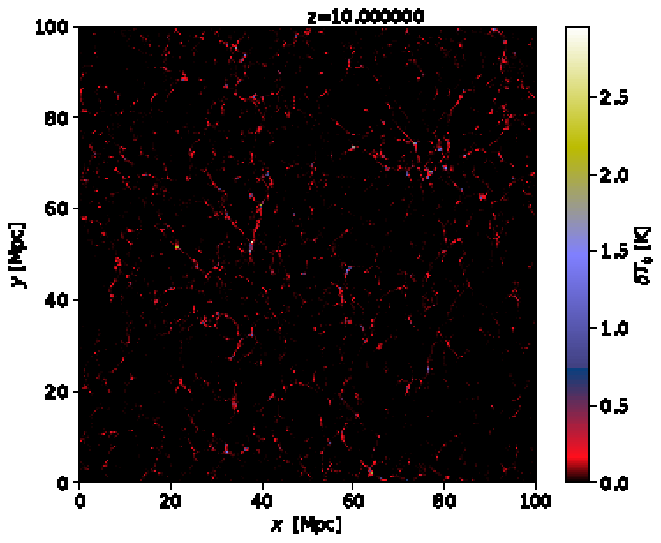}
%\hspace{0.1cm}
\epsfxsize=2.in\epsfbox{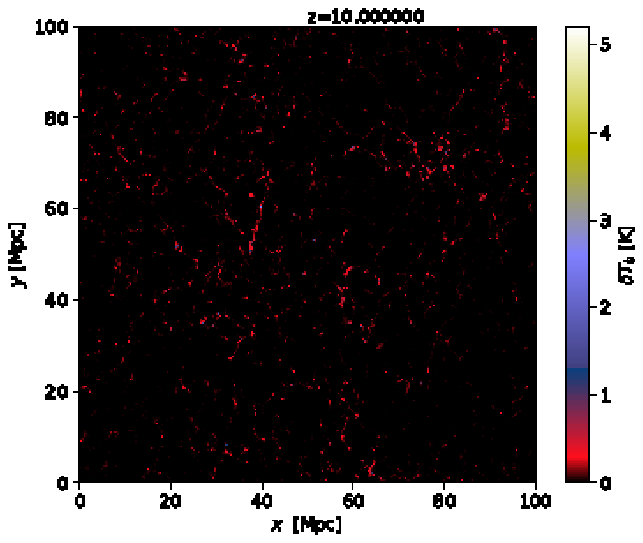}
}
\caption{The 21 cm line signal at $z=32,20,16,10$ with the highest redshift shown in the first row and the lowest redshift in the last row. In each panel from {\sl left} to {\sl right} are shown
the simulations for $B_0=0$ ({\sl far left}) and subsequently for the cases
$B_0=5$ nG and the spectral magnetic indices $n_B=-1.5$, $n_B=-2.2$ and $n_B=-2.9$.
} 
\label{figTb}
\end{figure}

The evolution of the ionization fraction of hydrogen is a key ingredient to determine the 21cm line signal. 
In figure \ref{figTb} the simulation boxes of the 21cm line signal are shown for redshifts $z=32$ to $z=10$ for the standard $\Lambda$CDM model and in the presence of the stochastic magnetic field. At $z=32$ hydrogen is neutral. At this epoch the 21 cm line signal is saturated and observed in emission, $\delta T_b>0$. This is an effect of not taking into account the details of Ly$\alpha$ coupling but rather  assuming that the spin temperature is much larger than the temperature of the CMB. The spatial distribution of the $\delta T_b$ traces the underlying matter density field. This can be appreciated when comparing the upper panels of
figures \ref{fig3} and \ref{figTb} which correspond to redshift $z=32$.

\begin{figure}[ht]
\centerline{
\epsfxsize=3.0in\epsfbox{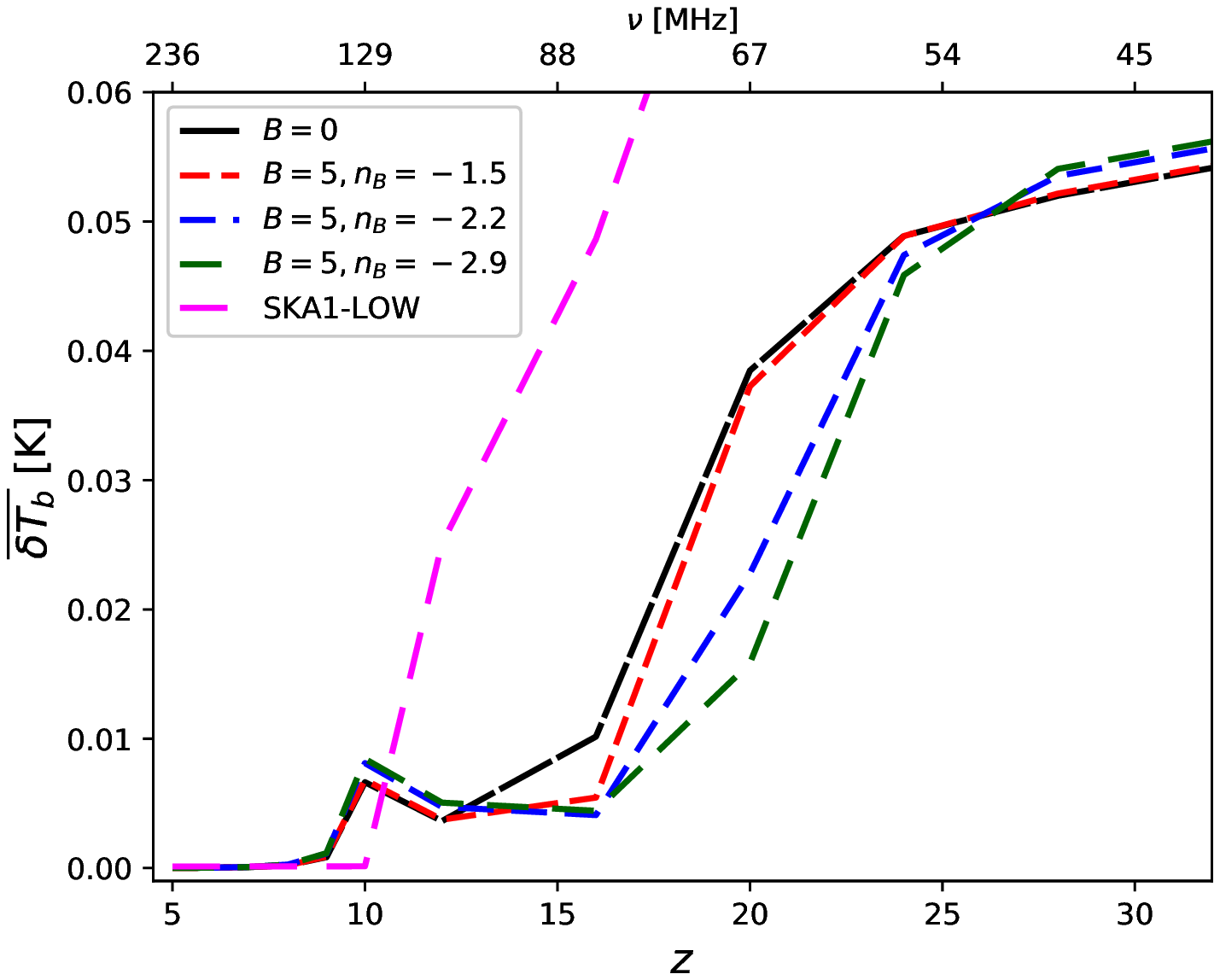}
}
\caption{The average 21 cm line signal as a function of redshift including sensitivity for SKA1-LOW.
} 
\label{figt21av}
\end{figure}

In figure \ref{figt21av} the average 21 cm line signal is shown as a function of redshift together with the projected sensitivity of SKA1-LOW of the Square Kilometre Array (SKA) \cite{Koopmans:2015sua}.
The baseline design of SKA1 will cover a frequency range of 50-350 MHz. In calculating the sensitivity 
we assumed one beam of bandwidth 300 MHz and an integration time of 1000h.

\begin{figure}[ht]
\centerline{
\epsfxsize=3.0in\epsfbox{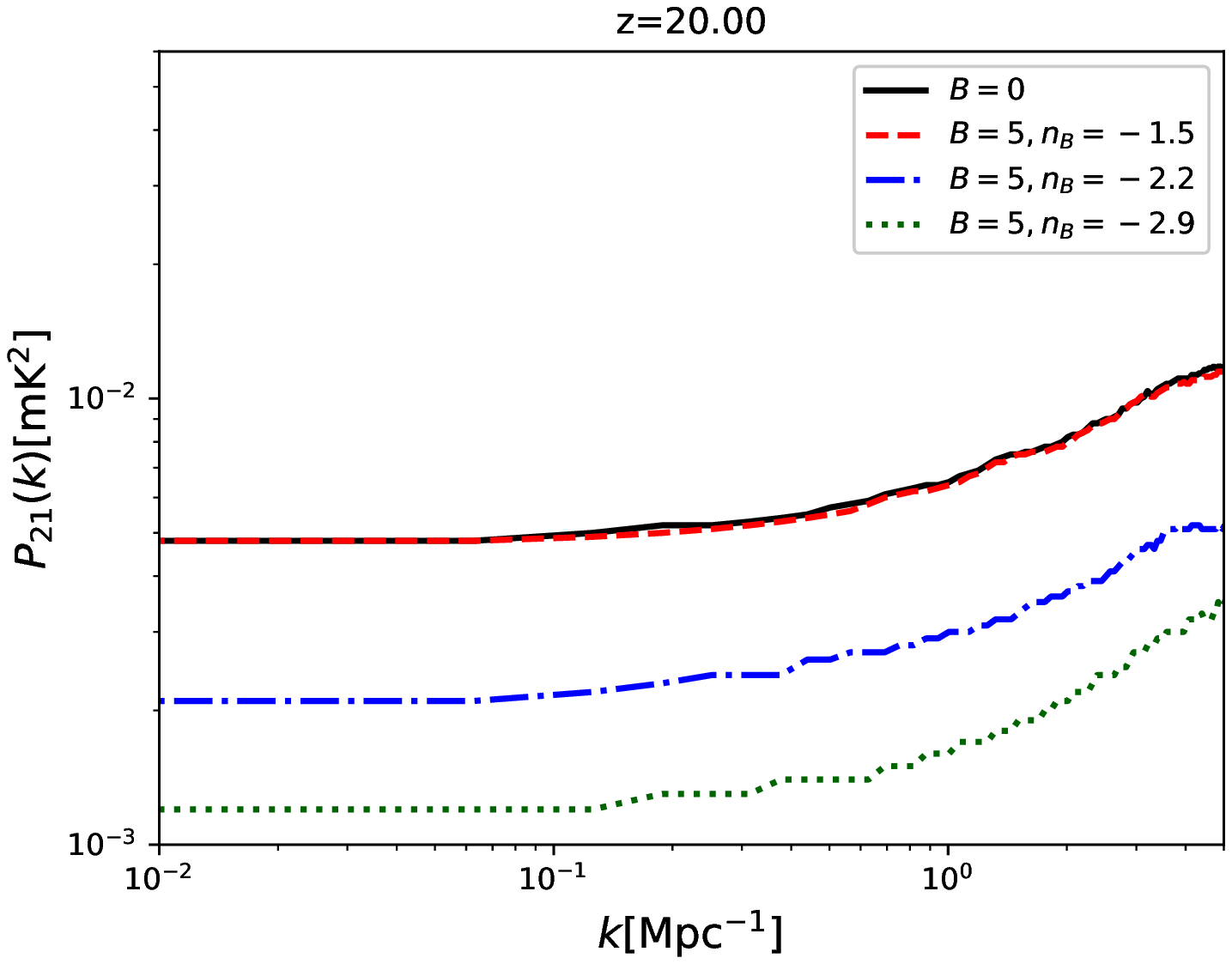}
\epsfxsize=3.0in\epsfbox{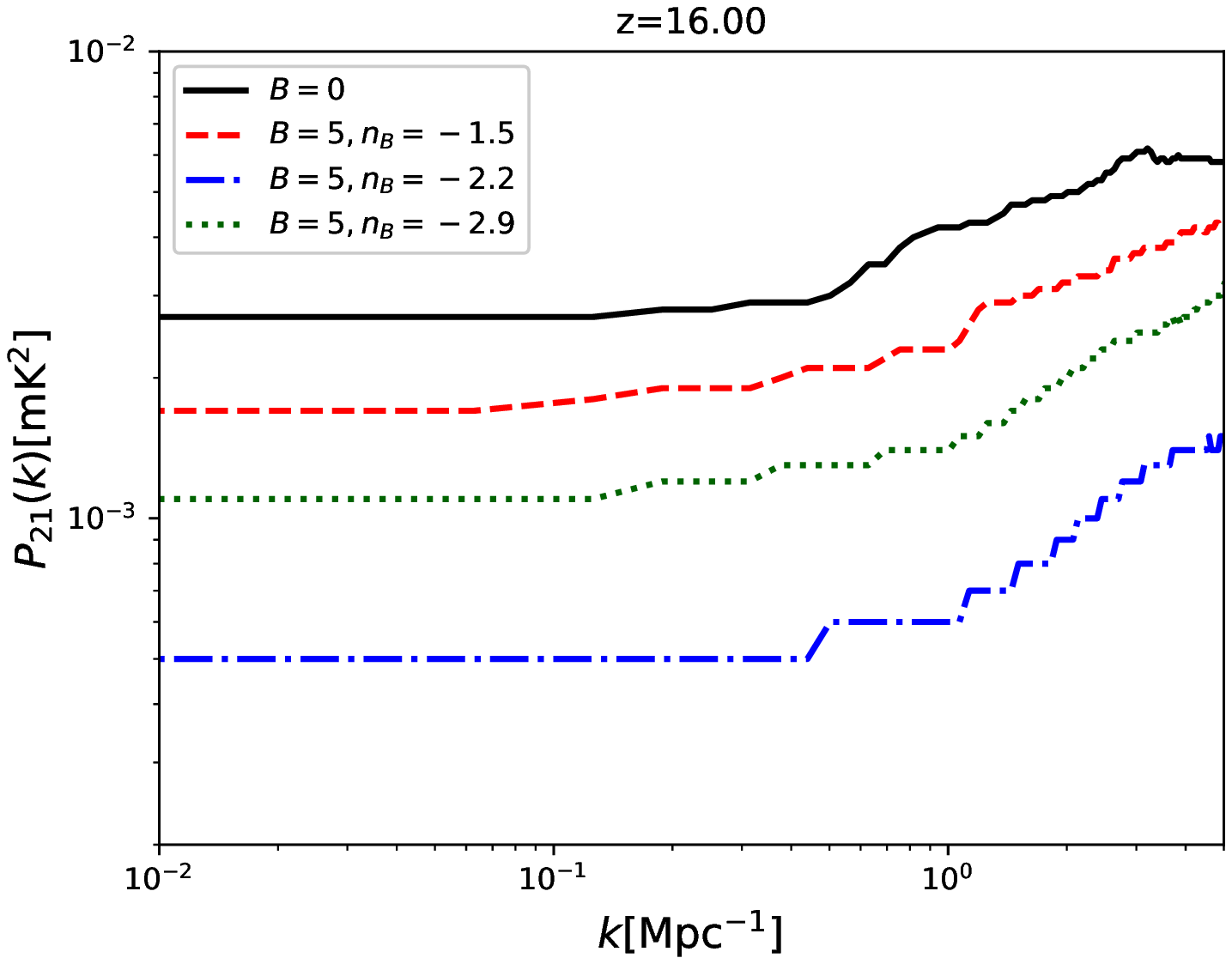}}
\caption{The estimated power spectrum of the 21 cm line signal $P_{21}(k)$ at redshifts
$z=20$ ({\sl  left}) and $z=16$ ({\sl right}). 
} 
\label{figPS21-z}
\end{figure}

\begin{figure}[ht]
\centerline{
\epsfxsize=3.0in\epsfbox{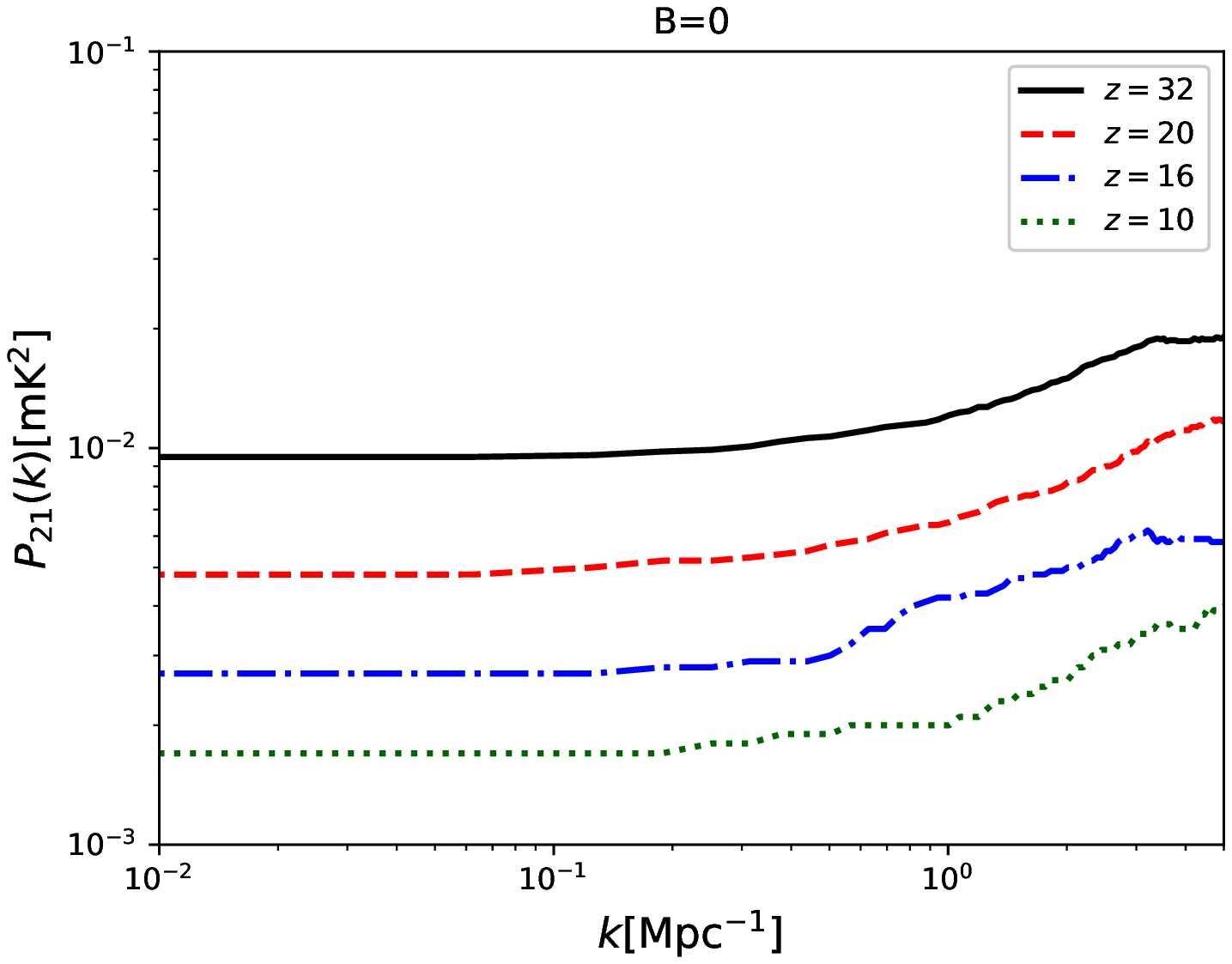}
\hspace{0.1cm}
\epsfxsize=3.0in\epsfbox{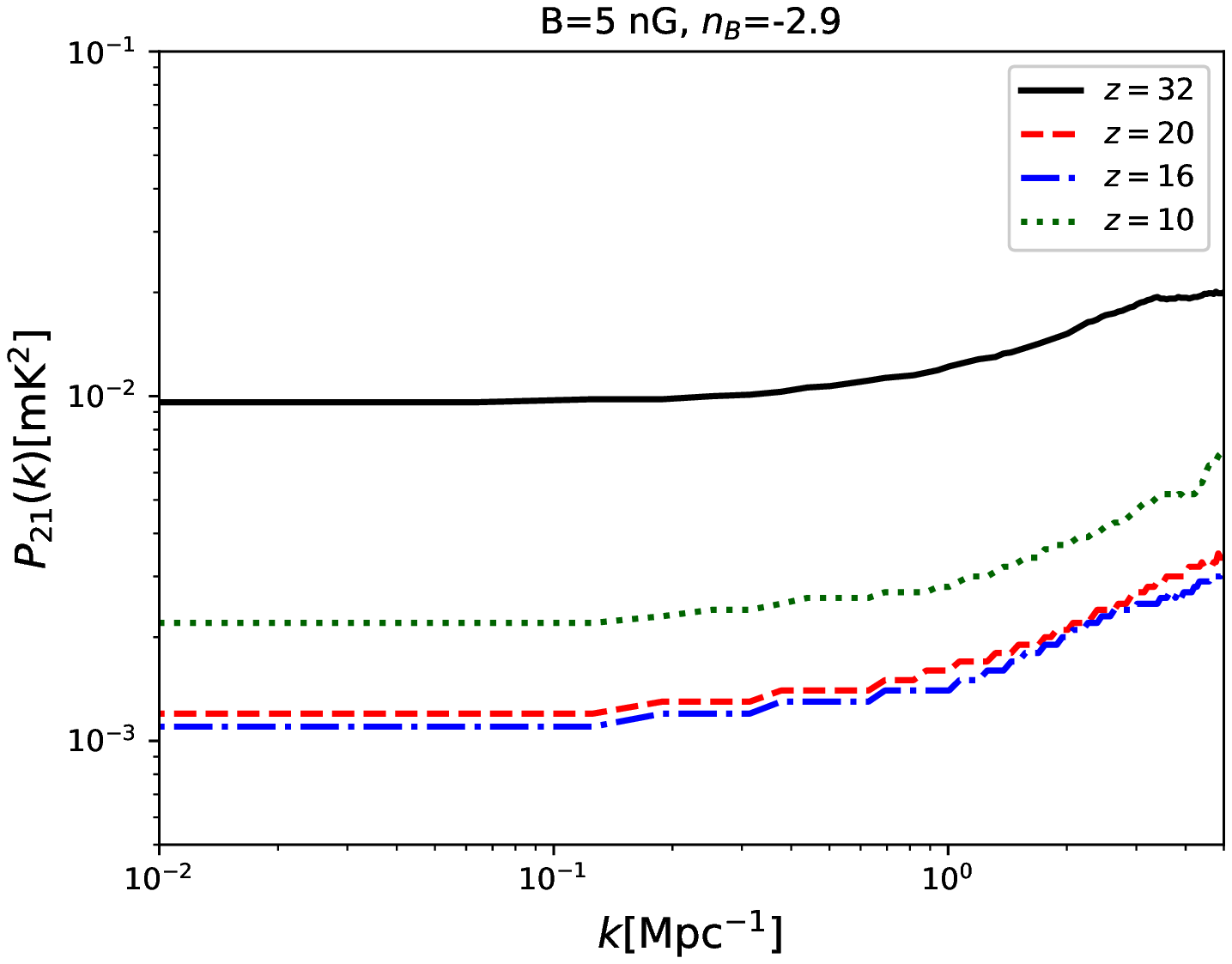}}
\caption{The estimated power spectrum of the 21 cm line signal $P_{21}(k)$ comparing all magnetic field models at different redshifts,
$B=0$ ({\sl left}) and $B=5$ nG, $n_B=-2.9$ ({\sl right}).
} 
\label{figPS21-B}
\end{figure}

In figure \ref{figPS21-z} the power spectra of the change in the CMB brightness temperature $P_{21}(k)$ for all magnetic fields models at different redshifts are shown for our simulations.
It is interesting to note that the feature introduced by the magnetic mode into the total linear matter power spectrum has left a mark on the power spectrum of the 21 cm line signal. Comparing curves for the magnetic spectral indices $n_B=-2.2$ and $n_B=-2.9$ at $z=20$ shows that whereas the former reaches a local maximum the latter rises steadily for large values of $k$. The difference in amplitude of the power spectra reflects the earlier beginning of EoR for smaller spectral indices, resulting in a suppression of the 21 cm line signal. This is also observed in the average signal $\overline{\delta T}_b$ in figure \ref{figt21av}.
In figure \ref{figPS21-B} the evolution with redshift of the power spectra of the change in the CMB brightness temperature $P_{21}(k)$ is shown for the two extreme cases, no magnetic mode, $B=0$, and a magnetic mode with $B=5$ nG and $n_B=-2.9$. 
Whereas the change in amplitude is the dominant feature, the change in spectral index of $P_{21}(k)$ is subleading in the evolution with $z$.

\section{Conclusions }
\label{s4}
\setcounter{equation}{0}

Primordial magnetic fields present since before decoupling add additional power on small scales to
the linear matter power spectrum. 
In using the modified linear matter power spectrum as initial condition for the simulation of the 
nonlinear density field clearly shows this effect. Moreover it subsequently changes the distribution of ionized hydrogen as well as the distribution of the 21 cm line signal. Simulations have been reported for 
magnetic fields of $B=5$ nG and different magnetic field indices, $n_B=-2.9$, $n_B=-2.2$ and $n_B=-1.5$ with the largest, visible effect for $n_B=-2.9$. Simulations have been run using the {\tt Simfast21} code.

When comparing the average 21 cm line signal with the projected sensitivity of the planned SKA1-LOW 
indicates that observations at frequencies above 120 MHz will be able to constrain parameters of
a primordial magnetic fields.

\section{Acknowlegements}

Financial support by Spanish Science Ministry grant FPA2015-64041-C2-2-P (FEDER) 
is gratefully acknowledged.

%%%%%%%%%%%%%%%%%%%%%%%%%%%%%%%%%%%%%%%%%%%%%%%%%%%%%%%%%%%%%%
%%%%%%%%%%%%%%%%%%%%%%%%%%%%%%%%%%%%%%%%%%%%%%%%%%%%%%%%%%%%%%

\bibliography{references}

\bibliographystyle{apsrev}

\end{document}